\documentclass[iop]{emulateapj}

\usepackage{amssymb}

\def\ts     {\thinspace}
\def\kms    {\ts km\ts s$^{-1}$}
\def\etal   {{\rm et\ts al.}}
\def\msol   {$M_{\odot}$}
\def\lsol   {$L_{\odot}$}

\def\oh     {{\rm OH}($^2\Pi_{1/2}\,J$=3/2$\to$1/2)}
\def\cii    {[C{\scriptsize II}]($^2P_{3/2}$$\to$$^2P_{1/2}$)}
\def\aco    {{\rm CO}($J$=1$\to$0)}

\def\bco    {{\rm CO}($J$=2$\to$1)}

\def\eco    {{\rm CO}($J$=5$\to$4)}
\def\fco    {{\rm CO}($J$=6$\to$5)}

\def\pco    {{\rm CO}($J$=16$\to$15)}

\slugcomment{draft version \today, accepted for publication in the Astrophysical Journal}

\shorttitle{ALMA imaging of [C{\scriptsize II}], OH, and far-infrared emission in a $z$=5.3 galaxy protocluster}
\shortauthors{Riechers et al.}

\begin{document}

\title{ALMA Imaging of Gas and Dust in a Galaxy Protocluster at Redshift 5.3:\ [C{\scriptsize II}] Emission in ``Typical'' Galaxies and Dusty Starbursts $\approx$1 Billion Years after the Big Bang}

\author{Dominik A.\ Riechers\altaffilmark{1,2}, Christopher L.\ Carilli\altaffilmark{3}, Peter L.\ Capak\altaffilmark{4}, Nicholas Z.\ Scoville\altaffilmark{2}, Vernesa Smol\v{c}i\'c\altaffilmark{5}, Eva Schinnerer\altaffilmark{6}, Min Yun\altaffilmark{7}, Pierre Cox\altaffilmark{8}, Frank Bertoldi\altaffilmark{9}, Alexander Karim\altaffilmark{9}, and Lin Yan\altaffilmark{4}}

\altaffiltext{1}{Astronomy Department, Cornell University, 220 Space Sciences Building, Ithaca, NY 14853, USA; dr@astro.cornell.edu}

\altaffiltext{2}{Astronomy Department, California Institute of
  Technology, MC 249-17, 1200 East California Boulevard, Pasadena, CA
  91125, USA}

\altaffiltext{3}{National Radio Astronomy Observatory, PO Box O, Socorro, NM 87801, USA}

\altaffiltext{4}{Spitzer Science Center, California Institute of Technology, MC 220-6, 1200 East California Boulevard, Pasadena, CA 91125, USA}

\altaffiltext{5}{University of Zagreb, Physics Department, Bijeni\v{c}ka cesta 32, 10002 Zagreb, Croatia}

\altaffiltext{6}{Max-Planck-Institut f\"ur Astronomie, K\"onigstuhl 17, D-69117 Heidelberg, Germany}

\altaffiltext{7}{Department of Astronomy, University of Massachusetts, Amherst, MA 01003, USA}

\altaffiltext{8}{ALMA Santiago Central Office, Alonso de Cordova 3107, Vitacura, Santiago, Chile}

\altaffiltext{9}{Argelander-Institut f\"ur Astronomie, Universit\"at
  Bonn, Auf dem H\"ugel 71, Bonn, D-53121, Germany}


\begin{abstract}

  We report interferometric imaging of \cii\ and \oh\ emission toward
  the center of the galaxy protocluster associated with the $z$=5.3
  submillimeter galaxy (SMG) AzTEC-3, using the Atacama Large
  (sub)Millimeter Array (ALMA). We detect strong [C{\scriptsize II}],
  OH, and rest-frame 157.7\,$\mu$m continuum emission toward the
  SMG. The \cii\ emission is distributed over a scale of 3.9\,kpc,
  implying a dynamical mass of 9.7$\times$10$^{10}$\,\msol, and a star
  formation rate (SFR) surface density of $\Sigma_{\rm
  SFR}$=530\,\msol\,yr$^{-1}$\,kpc$^{-2}$.  This suggests that AzTEC-3
  forms stars at $\Sigma_{\rm SFR}$ approaching the Eddington limit
  for radiation pressure supported disks. We find that the OH emission
  is slightly blueshifted relative to the [C{\scriptsize II}] line,
  which may indicate a molecular outflow associated with the peak
  phase of the starburst. We also detect and dynamically resolve \cii\
  emission over a scale of 7.5\,kpc toward a triplet of Lyman-break
  galaxies with moderate UV-based SFRs in the protocluster at
  $\sim$95\,kpc projected distance from the SMG. These galaxies are
  not detected in the continuum, suggesting far-infrared SFRs of
  $<$18--54\,\msol\,yr$^{-1}$, consistent with a UV-based estimate of
  22\,\msol\,yr$^{-1}$.  The spectral energy distribution of these
  galaxies is inconsistent with nearby spiral and starburst galaxies,
  but resembles those of dwarf galaxies. This is consistent with
  expectations for young starbursts without significant older stellar
  populations. This suggests that these galaxies are significantly
  metal-enriched, but not heavily dust-obscured, ``normal''
  star-forming galaxies at $z$$>$5, showing that ALMA can detect the
  interstellar medium in ``typical'' galaxies in the very early
  universe.

\end{abstract}


\section{Introduction}

The first massive galaxies in the universe are expected to rapidly
grow in the most massive dark matter halos at early cosmic epochs
(e.g., Efstathiou \& Rees \citeyear{er88}; Kauffmann et
al.\ \citeyear{kau99}).  Such high overdensities in the dark matter
distribution are expected to be associated accordingly with
overdensities of baryonic matter, and thus, protoclusters of galaxies
(e.g., Springel et al.\ \citeyear{spr05}). The bulk of the stellar
mass in the most massive galaxies in these halos likely grows in
short, episodic bursts associated with major gas-rich mergers and/or
peak phases of gas accretion from the intergalactic medium (e.g.,
Blain et al.\ \citeyear{bla04}).  These starbursts, in turn, may
significantly enrich the galaxy's environment with heavy elements
through winds and outflows (e.g., McKee \& Ostriker \citeyear{mo77};
Sturm et al.\ \citeyear{stu11}; Spoon et al.\ \citeyear{spo13}).

The identification of massive starburst galaxies at the highest
redshifts may be the most promising way to find such exceptional
cosmic environments. The most intense starbursts are commonly
enshrouded by dust, rendering them difficult to identify at rest-frame
UV/optical wavelengths (e.g., Smail et al.\ \citeyear{sma97}; Hughes
et al.\ \citeyear{hug98}; Chapman et al.\ \citeyear{cha03}). The
dust-absorbed stellar light is re-emitted at rest-frame far-infrared
(FIR) wavelengths, making such galaxies bright in the observed-frame
(sub-)millimeter at high redshift (so-called submillimeter galaxies,
or SMGs; see review by Blain et al.\ \citeyear{bla02}).

\begin{figure*}
\epsscale{1.15}
\plotone{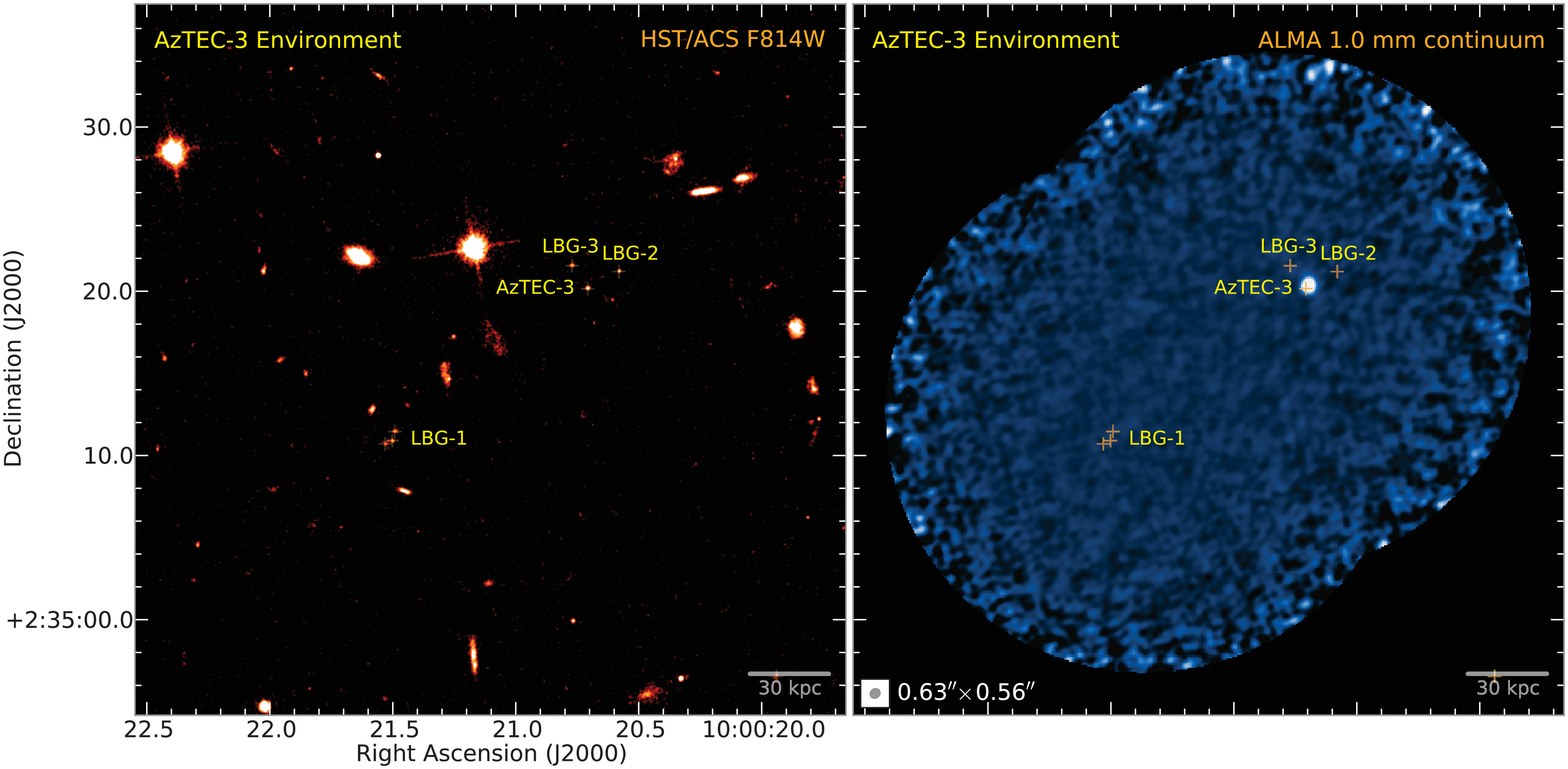}
\vspace{-2mm}

\caption{{\em HST}/ACS F814W ({\em left}) and ALMA 1.0\,mm continuum
  image (rest-frame 157.7\,$\mu$m; {\em right}) of the targeted
  region.  Two pointings were observed to cover AzTEC-3 at $z$=5.3 and
  five candidate companion Lyman-break galaxies (positions are
  indicated by plus signs; LBG-1 contains three components).  The
  1.0\,mm continuum image was obtained by averaging the three
  [C{\scriptsize II}] line-free spectral windows (corrected for
  primary beam attenuation).  The rms at the phase centers is
  $\sim$58\,$\mu$Jy\,beam$^{-1}$, and increases outwards due to the
  primary beam response. The synthesized beam size of
  0.63\,$''$$\times$0.56\,$''$ is indicated in the bottom left corner
  of the {\em right} panel.  \label{f1}}
%
\end{figure*}

We have recently identified AzTEC-3, a gas-rich SMG at $z$=5.3
(Riechers et al.\ \citeyear{rie10}, hereafter R10; Capak et
al.\ \citeyear{cap11}, hereafter C11) in the AzTEC 1.1\,mm study of
the Cosmic Evolution Survey (COSMOS) field (Scoville et
al.\ \citeyear{sco07}; Scott et al.\ \citeyear{sco08}). AzTEC-3 has a
relatively compact ($<$4\,kpc radius), highly excited molecular gas
reservoir of 5.3$\times$10$^{10}$\,\msol\ (determined through the
detection of three CO lines; R10), which gets converted into stars at
a rate of $>$1000\,\msol\,yr$^{-1}$ (C11). Its current stellar mass
is estimated to be
$M_\star$=(1.0$\pm$0.2)$\times$10$^{10}$\,\msol\ (C11).\footnote{These
  estimates depend on the assumed stellar initial mass function, see,
  e.g., Dwek et al.\ (\citeyear{dwe11}).}

The environment of AzTEC-3 represents one of the most compelling
pieces of observational evidence for the hierarchical picture of
massive galaxy evolution. The massive starburst galaxy is associated
with a $>$11-fold overdense structure of ``normal'' star-forming
galaxies at the same redshift\footnote{Based on photometric redshifts
  and several spectroscopic confirmations (C11).} that extends out to
$>$13\,Mpc on the sky, with $>$10 galaxies within the central
(co-moving) $\sim$2\,Mpc radius region (C11). The protocluster
galaxies alone (including the SMG) place a lower limit of
4$\times$10$^{11}$\,\msol\ on the mass of dark and luminous matter
associated with this region (C11). However, our current understanding
of this exceptional cosmic environment is dominantly based on the
rest-frame UV/optical properties of all galaxies except the SMG, and
thus, may be incomplete due to lacking information on the gas and dust
in their interstellar media (ISM).

Here we report 158\,$\mu$m \cii, 163\,$\mu$m \oh, and rest-frame
157.7\,$\mu$m dust continuum imaging toward the center of the galaxy
protocluster associated with the $z$=5.3 SMG AzTEC-3 with ALMA. The
\cii\ line is the dominant cooling line of the cold\footnote{Cold here
  means $\ll$10$^4$\,K, i.e., in the regime where dust cooling through
  (far-)infrared emission is prevalent, and below the regime where
  cooling through hydrogen lines dominates.} ISM in star-forming
galaxies (where it can carry up to 1\% of $L_{\rm FIR}$; e.g., Israel
et al.\ \citeyear{isr96}), typically much brighter than CO lines, and
traces regions of active star formation (photon-dominated regions, or
PDRs) and the cold, neutral atomic medium (CNM; e.g., Stacey et
al.\ \citeyear{sta91}). It thus is an ideal tracer for the
distribution, dynamics, and enrichment of the ISM out to the most
distant galaxies, but it was only detected in some of the most
luminous quasars and starburst galaxies in the past (e.g., Maiolino et
al.\ \citeyear{mai05}, \citeyear{mai09}; Walter et
al.\ \citeyear{wal09}; Stacey et al.\ \citeyear{sta10}; Wagg et
al.\ \citeyear{wag10}; Valtchanov et al.\ \citeyear{val11}; Riechers
et al.\ \citeyear{rie13}; Wang et al.\ \citeyear{wan13}) -- i.e.,
systems that are much more extreme than typical protocluster
galaxies. Previous searches for [C{\scriptsize II}] emission in
typical and/or ultraviolet-luminous galaxies at $z$$>$5 have been
unsuccessful (e.g., Walter et al.\ \citeyear{wal12}; Kanekar et
al.\ \citeyear{kan13}; Ouchi et al.\ \citeyear{ouc13}; Gonzalez-Lopez
et al.\ \citeyear{gl14}), and it is important to understand what role
environment may play for the detectability of such objects. The
far-infrared lines of the OH radical are important for the H$_2$O
chemistry and cooling budget of star-forming regions, and they are
critical tracers of molecular outflows (e.g., Sturm et
al.\ \citeyear{stu11}; Gonzalez-Alfonso et al.\ \citeyear{ga12}), but
OH was only detected in a single galaxy at cosmological distances to
date (Riechers et al.\ \citeyear{rie13}).  We use a concordance, flat
$\Lambda$CDM cosmology throughout, with $H_0$=71\,\kms\,Mpc$^{-1}$,
$\Omega_{\rm M}$=0.27, and $\Omega_{\Lambda}$=0.73 (Spergel
\etal\ \citeyear{spe03}, \citeyear{spe07}).

\section{Observations}

\begin{figure*}
\epsscale{1.15}
\plotone{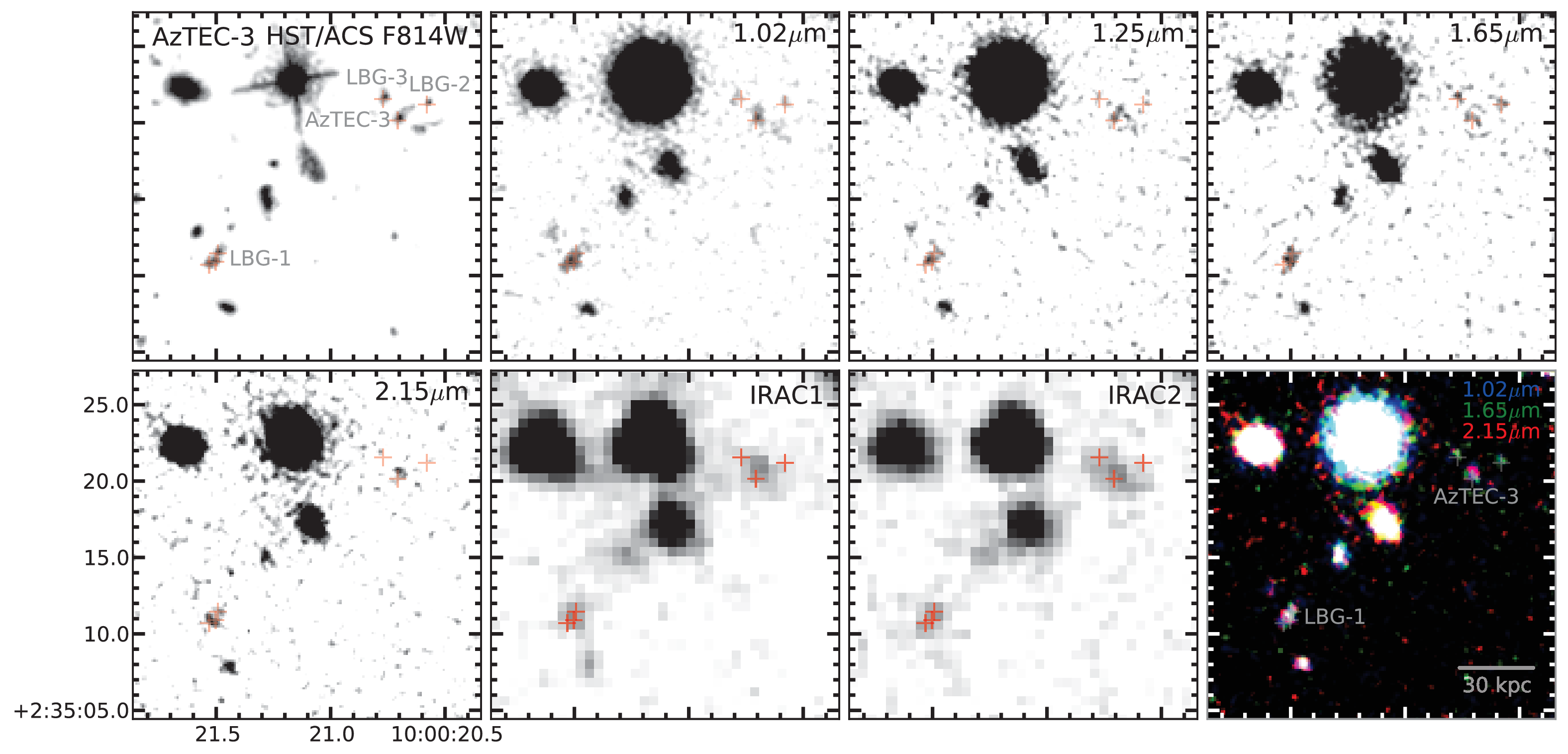}
\vspace{-2mm}

\caption{Images of the field observed with ALMA in the {\em HST}/ACS
  $i$ (F814W), UltraVISTA $Y$, $J$, $H$, $Ks$ (1.02, 1.25, 1.65,
  2.15\,$\mu$m), {\em Spitzer} IRAC1 (3.6\,$\mu$m), and IRAC2
  (4.5\,$\mu$m) bands, and $YHKs$ three-color image (Scoville et
  al.\ \citeyear{sco07}; McCracken et al.\ \citeyear{mcc12}; Sanders
  et al.\ \citeyear{san07}). Adaptive smoothing has been applied to
  the {\em HST}/ACS image. The plus signs indicate the same positions
  as in Fig.~\ref{f1}.
  \label{f10}}
%
\end{figure*}

We observed the \cii\ transition line ($\nu_{\rm rest} =
1900.5369\,$GHz, redshifted to 301.72\,GHz, or 994\,$\mu$m, at
$z$=5.299), using ALMA.  Observations were carried out with 16--24
usable 12\,m antennas under good 870\,$\mu$m weather conditions
(precipitable water vapor of 0.64--1.76\,mm) for 4\,tracks in the
cycle 0 extended configuration (longest baseline:\ 402\,m) between
2012 April 11 and May 17, and for 1\,track in the cycle 0 compact
configuration\footnote{Given the increased number of antennas
available, this configuration also contained antennas on longer
baselines than the originally advertised 125\,m.} (shortest baseline:\
21\,m) on 2012 November 18 (three additional tracks were discarded due
to poor data quality). This resulted in 125\,min on-source time, which
was evenly split over two (slightly overlapping) pointings (primary
beam FWHM diameter at 994\,$\mu$m:\ 20$''$). The nearby radio quasar
J1058+015 was observed regularly for pointing, bandpass, amplitude and
phase calibration. Fluxes were derived relative to Titan or
Callisto. We estimate the overall accuracy of the calibration to be
accurate within $\sim$20\%.

The correlator was set up to target two spectral windows of 1.875\,GHz
bandwidth each at 0.488\,MHz (0.48\,\kms ) resolution (dual
polarization) in each sideband. The [C{\scriptsize II}] line was
centered in one spectral window in the upper sideband. The other three
windows were used to measure the continuum emission $\sim$2\,GHz above
and $\sim$10--12\,GHz below the line frequency. This setup also
covered the 163\,$\mu$m \oh\ line (components at $\nu_{\rm rest} =
1834.74735$ and 1837.81682\,GHz, redshifted to 291.27597 and
291.76327\,GHz at $z$=5.299),\footnote{These components are due to a
  [$P$=$(-)$$\to$$(+)$, $P$=$(+)$$\to$$(-)$] $\Lambda$-doublet in the
  orbital angular momentum of the OH radical.  Each $P$ component of
  the $J$=3/2$\to$1/2 transition consists of an unresolved $F$ triplet
  of transitions due to hyperfine structure splitting of the levels
  (see, e.g., Genzel et al.\ \citeyear{gen85}; Cesaroni \&
  Walmsley~\citeyear{cw91}).}  as well as the \pco\ line ($\nu_{\rm
  rest} = 1841.345506\,$GHz, redshifted to 292.32346\,GHz), in the
lower sideband.

The Astronomical Image Processing System (AIPS) and Common Astronomy
Software Applications (CASA) packages were used independently for data
reduction and analysis to better quantify potential uncertainties in
the calibration. Primary beam-corrected mosaics were created using the
FLATN task in AIPS. All data were mapped using the CLEAN algorithm
with ``natural'' weighting; resulting in a synthesized beam size of
0.63\,$''$$\times$0.55\,$''$ at the redshifted [C{\scriptsize II}]
line frequencies (0.63\,$''$$\times$0.56\,$''$ when averaging over the
three [C{\scriptsize II}] line-free spectral windows). The final rms
noise when averaging over all spectral windows (i.e., over a total of
7.5\,GHz of bandwidth) is $\sim$50\,$\mu$Jy\,beam$^{-1}$ in the phase
centers of both pointings, and scales as expected for thermal noise in
narrower frequency bins (see figure captions).\footnote{A narrow,
  limited frequency range at 301.8\,GHz (corresponding to
  approximately --70\,\kms\ on the velocity scale used throughout the
  paper) is affected by a narrow atmospheric feature, which causes the
  the sensitivity to be $\lesssim$15\%--20\% worse than nominal. The
  feature is narrow compared to all detected emission lines. Also, it
  is offset from virtually all line emission in LBG-1, and from the
  center of the [C{\scriptsize II}] line in AzTEC-3. This feature may
  be responsible for the slightly higher apparent rms noise level in
  the rightmost panel of Figure \ref{f5}. Other known atmospheric
  features in the frequency range of our observations are very weak,
  and affect the sensitivity at few per cent level at most, and all
  lie outside the spectral windows that contain contain the
  [C{\scriptsize II}] and OH emission.} This sensitivity level is
consistent with standard theoretical estimates when assuming the
atmospheric conditions and instrument configurations used for our
observations.

\section{Results}

\begin{figure*}
\epsscale{1.15}
\plotone{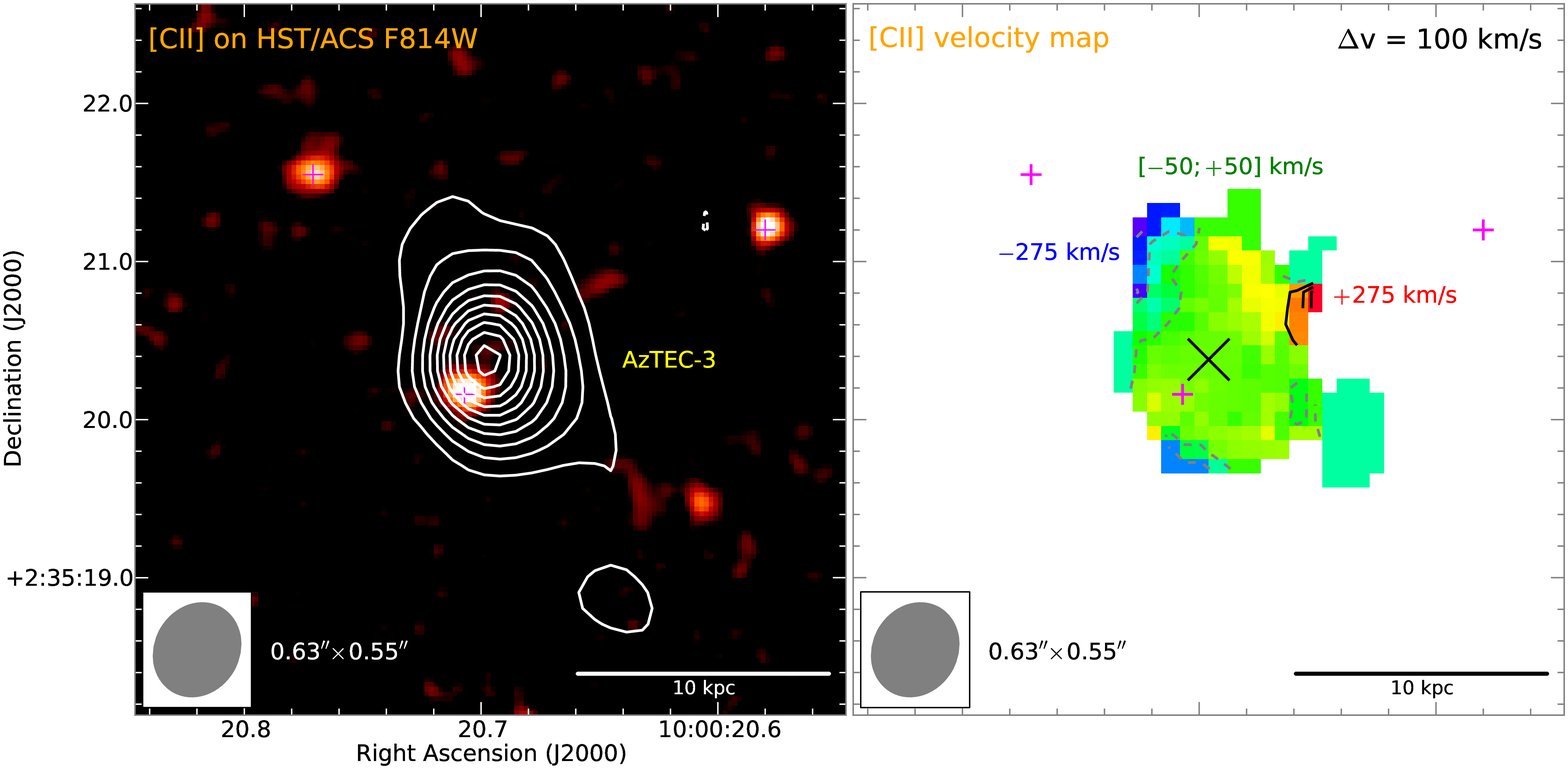}
\vspace{-2mm}

\caption{Velocity-integrated ALMA \cii\ contours overlaid on {\em
    HST}/ACS F814W image ({\em left}), and \cii\ velocity map ({\em
    right}) toward AzTEC-3. Continuum emission has been subtracted
    from all maps.  {\em Left:}\ The map is averaged over 468.75\,MHz
    (466\,\kms ). Contours start at $\pm$4$\sigma$ and are in steps of
    4$\sigma$ (1$\sigma$=200\,$\mu$Jy\,beam$^{-1}$ at the phase
    center).  The synthesized beam size of
    0.63\,$''$$\times$0.55\,$''$ is indicated in the bottom left
    corners.  The plus signs indicate the same positions as in
    Fig.~\ref{f1}.  The [C{\scriptsize II}] peak position (black cross
    in {\em right} panel) is consistent with those of the CO and FIR
    continuum emission (R10).  {\em Right:}\ First moment map (i.e.,
    intensity weighted mean velocity image) of the [C{\scriptsize II}]
    velocity structure. The colors indicate the velocity
    gradient. Contours are shown in steps of 100\,\kms, with dashed
    (solid) contours showing blueshifted (redshifted) emission
    relative to zero velocity.  The velocity scale is relative to
    $z$=5.2988.  \label{f2}}
%
\end{figure*}

\subsection{Continuum}

We have detected 1.0\,mm continuum emission toward the $z$=5.3 SMG
AzTEC-3 (Figure~\ref{f1}). From two-dimensional Gaussian fitting to a
data set averaged over the line-free spectral windows, we find a
continuum flux density of 6.20$\pm$0.25\,mJy (Table~\ref{t1}) and a
deconvolved continuum source size of
0.40$''$$^{+0.04''}_{-0.04''}$$\times$0.17$''$$^{+0.08''}_{-0.17''}$
(2.5$\times$1.1\,kpc$^2$ at $z$$\simeq$5.3;
Tab.~\ref{t2}).\footnote{Errors are determined as described by Condon
  (\citeyear{con97}).}

The two pointings also covered five Lyman-break galaxies with
photometric redshifts close to (or colors consistent with) $z$=5.3,
with COSMOS optical IDs 1447523, 1447526 (Ilbert et
al.\ \citeyear{ilb09}; ``LBG-2'' and ``LBG-3'' in the following;
$\sim$1.7$''$ north-east and $\sim$2.2$''$ north-west of AzTEC-3), and
1447524 (a triple of sources $\sim$15$''$ south-east of AzTEC-3,
targeted by the second pointing; ``LBG-1'' below and in Fig.~\ref{f1};
see C11 for more details on the source identifications). Besides color
information, LBG-1 has an optical spectroscopic redshift of $z_{\rm
spec}$=5.300 that is likely valid for all of its three components
(LBG-1a, LBG-1b, and LBG-1c below), and LBG-3 has a photometric
redshift of $z_{\rm phot}$=5.269 (C11). No 1.0\,mm continuum emission
is detected toward any of these galaxies, with 3$\sigma$ upper limits
of 0.24\,mJy\,beam$^{-1}$ for LBG-2 and LBG-3 (conservative limits are
quoted due to potential sidelobe residuals from AzTEC-3), and
0.45\,mJy for LBG-1 (or 0.15--0.24\,mJy\,beam$^{-1}$ for the three LBG
subcomponents when accounting for potential source overlap at the
current spatial resolution).  Based on their small sizes in {\em
HST}/ACS F814W imaging data (Fig.~\ref{f1}), we assume in the
following that LBG-2 and LBG-3 are spatially unresolved by our ALMA
observations. Given the spatial separation of the three components of
LBG-1 (Fig.~\ref{f1}) and the spatial extent of the \cii\ emission
(see below), we assume that the continuum emission in this source is
resolved over $\sim$3\,beams for the above estimate. We adopt this
spatially-integrated limit (rather than more sensitive limits for
individual or stacked LBG components) in the following, to enable
comparison with existing resolution-limited photometry at other
wavelengths (see, e.g., Fig.~\ref{f10}).

\begin{figure*}
\vspace{-23mm}
\epsscale{1.15}
\plotone{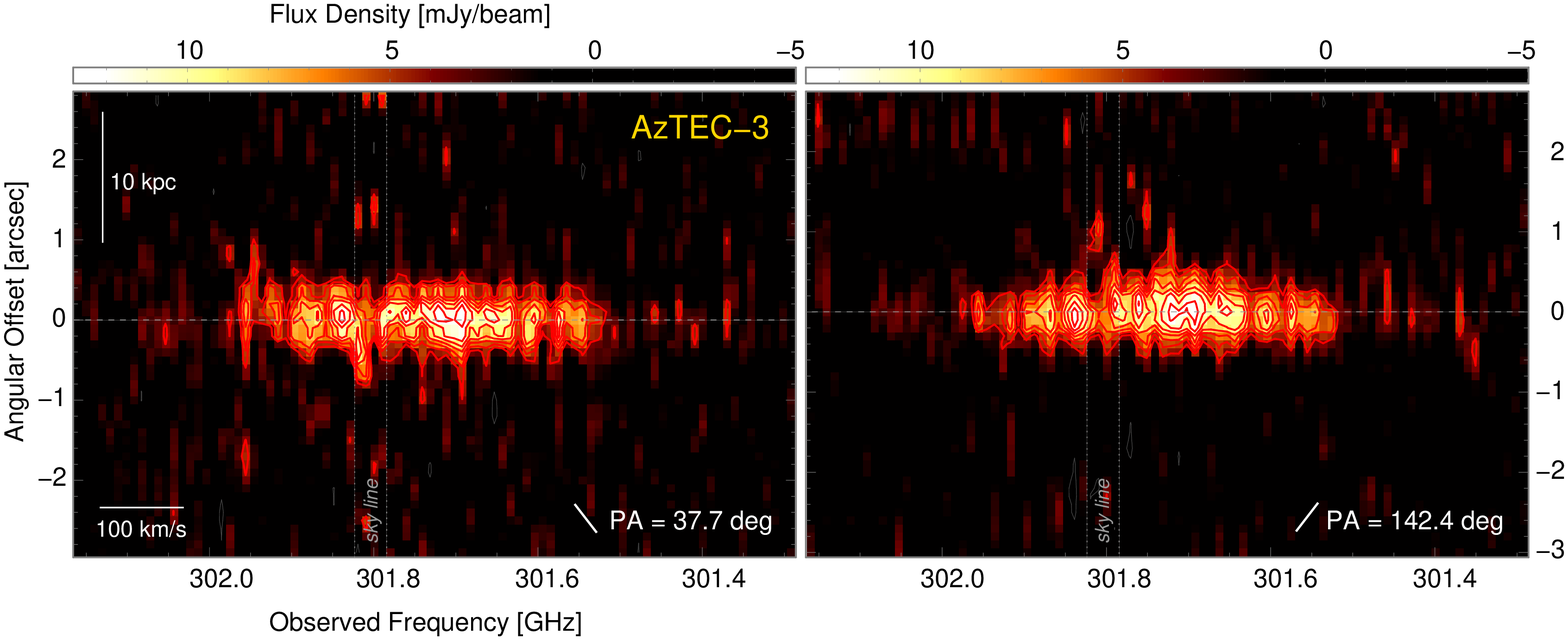}
\vspace{-23mm}

\caption{ALMA \cii\ position-velocity ($p-v$) diagrams of
  AzTEC-3. Continuum emission has been subtracted from the data
  cube. {\em Left:}\ $p-v$ diagram along an axis with a position angle
  of 37.7$^\circ$ (direction as indicated by the inclined, short solid
  line), which connects the line centroid with the peak to the
  north-east seen in the blue velocity range in Fig.~\ref{f4a}. {\em
    Right:}\ $p-v$ diagram along an axis perpendicular to that in the
  {\em left} panel (position angle of 142.4$^\circ$). Data are shown
  at a spectral resolution of 9.77\,MHz ($\sim$10\,\kms ). The spatial
  pixel scale is the same as in Fig.~\ref{f2}. Contours are shown in
  steps of 1$\sigma$, starting at $\pm$3$\sigma$. Negative contours
  are shown as thin gray lines. The dotted lines indicate a narrow
  frequency range where the sensitivity may be reduced by a few per
  cent to $\lesssim$15\%--20\% relative to the nominal value due to a
  weak atmospheric absorption feature (see discussion in
  Sect.~2).\label{f2a}}
\vspace{-5mm}

\end{figure*}


\begin{deluxetable*}{lcccccccc}
\tabletypesize{\scriptsize}
\tablecaption{[C{\scriptsize II}] and continuum properties of sources in the $z$=5.3 protocluster. \label{t1}}
\tablehead{
Target & redshift & $S_{\rm CII}$ & d$v_{\rm CII}$ & $I_{\rm CII}$ & $L'_{\rm CII}$               & $L_{\rm CII}$    & $S_{1.0{\rm mm}}$ & log$_{10}$($L_{\rm CII}$/$L_{\rm FIR}$) \\
       &          & [mJy]         & [\kms ]        & [Jy\,\kms ]   & [10$^{10}$\,K\,\kms\,pc$^2$] & [10$^9$\,\lsol ] & [mJy] & }
\startdata
AzTEC-3     & 5.2988$\pm$0.0001 & 18.4$\pm$0.5  & 421$\pm$19  & 8.21$\pm$0.29 & 3.05$\pm$0.11 & 6.69$\pm$0.23 & 6.20$\pm$0.25 & --3.40 \\
LBG-1       & 5.2950$\pm$0.0002 & 8.99$\pm$0.73 & 218$\pm$24  & 2.08$\pm$0.18 & 0.77$\pm$0.07 & 1.69$\pm$0.15 & $<$0.45       & $>$(--2.50) \\
            &                   &               &             &               &               &               & $<$0.15--0.24\tablenotemark{a,b} & $>$(--2.03)--(--2.22)\tablenotemark{c} \\
LBG-2 & & & & $<$0.21\tablenotemark{b,d} & $<$0.08\tablenotemark{b,d} & $<$0.17\tablenotemark{b,d} & $<$0.24\tablenotemark{b} & \\
LBG-3 & & & & $<$0.21\tablenotemark{b,d} & $<$0.08\tablenotemark{b,d} & $<$0.17\tablenotemark{b,d} & $<$0.24\tablenotemark{b} & \\
\vspace{-3mm}
\enddata
{\tablecomments{All quoted upper limits are 3$\sigma$.}}
\tablenotetext{a}{Individual limits for subcomponents LBG-1a, LBG-1b, and LBG-1c.}
\tablenotetext{b}{Point source limits in mJy\,beam$^{-1}$, Jy\,beam$^{-1}$\,\kms, or equivalent.}
\tablenotetext{c}{Assuming that the far-infrared continuum emission is not resolved by our observations.}
\tablenotetext{d}{Assuming the same line width and redshift as measured for LBG-1. Assuming an average or median of the line width of the subcomponents of LBG-1 would result in $\sim$10\%--15\% lower limits. We consider this difference in limits negligible compared to other sources of uncertainty.}
\end{deluxetable*}


\subsection{\cii\ Line Emission}

\begin{figure*}
\epsscale{1.15}
\plotone{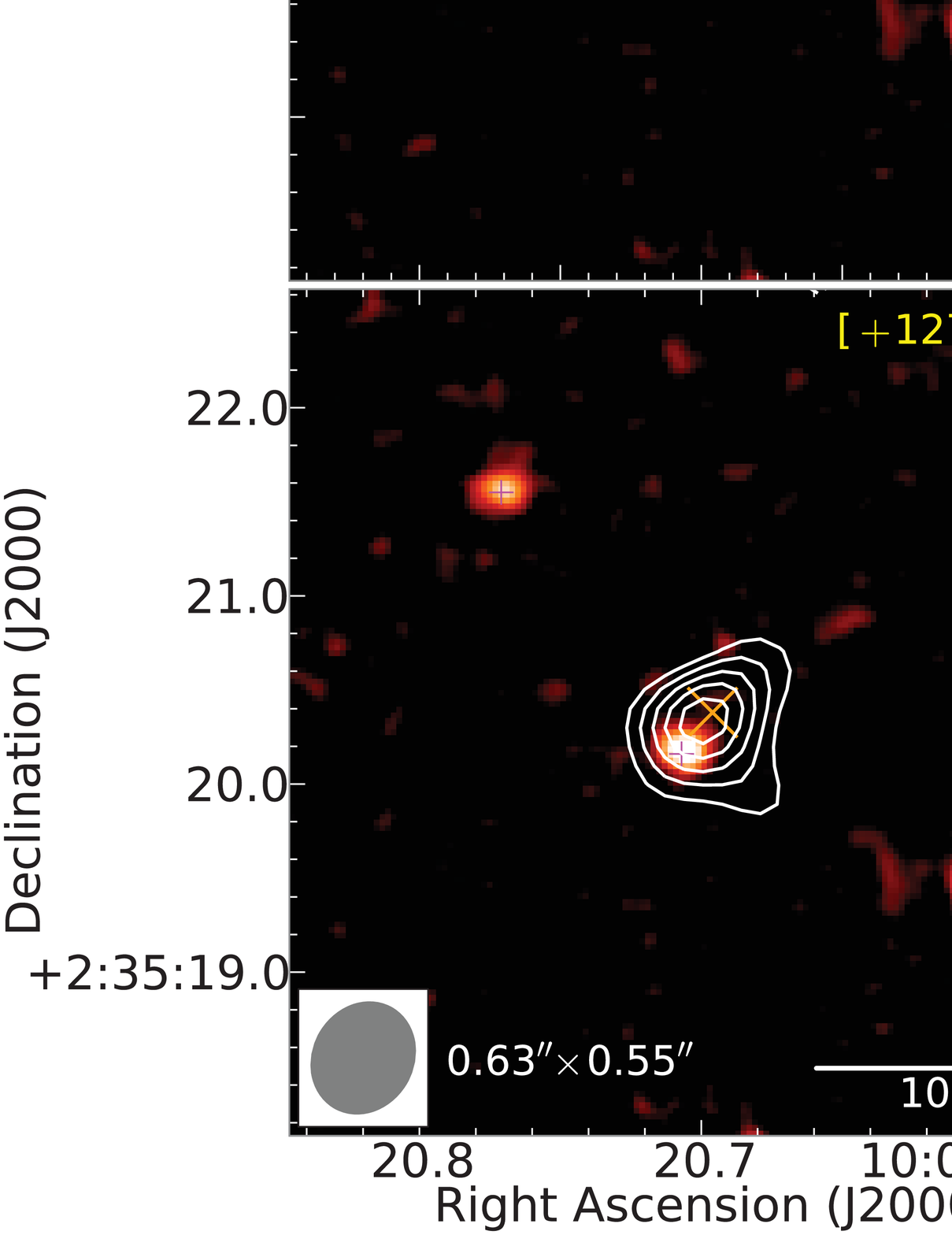}
\vspace{-2mm}

\caption{ALMA \cii\ velocity channel contours overlaid on {\em
    HST}/ACS F814W image toward AzTEC-3.  Velocity channels are
  averaged over $\sim$19.53\,MHz ($\sim$19\,\kms ).  Contours start at
  $\pm$3$\sigma$ and are in steps of 1$\sigma$
  (1$\sigma$=974\,$\mu$Jy\,beam$^{-1}$ at the phase center). The noise
  close to $-$70\,\kms\ is slightly higher due to a weak atmospheric
  absorption feature (see Sect.~2). Velocity ranges in \kms\ are
  indicated in the top right corner of each panel. The synthesized
  beam size is the same as in Fig.~\ref{f2}.  The velocity scale is
  the same as in Fig.~\ref{f2}. The crosses and plus signs indicate
  the same positions as in Fig.~\ref{f2}.  \label{f4}}
%
\end{figure*}

\begin{figure}[b]
\epsscale{1.17}
\plotone{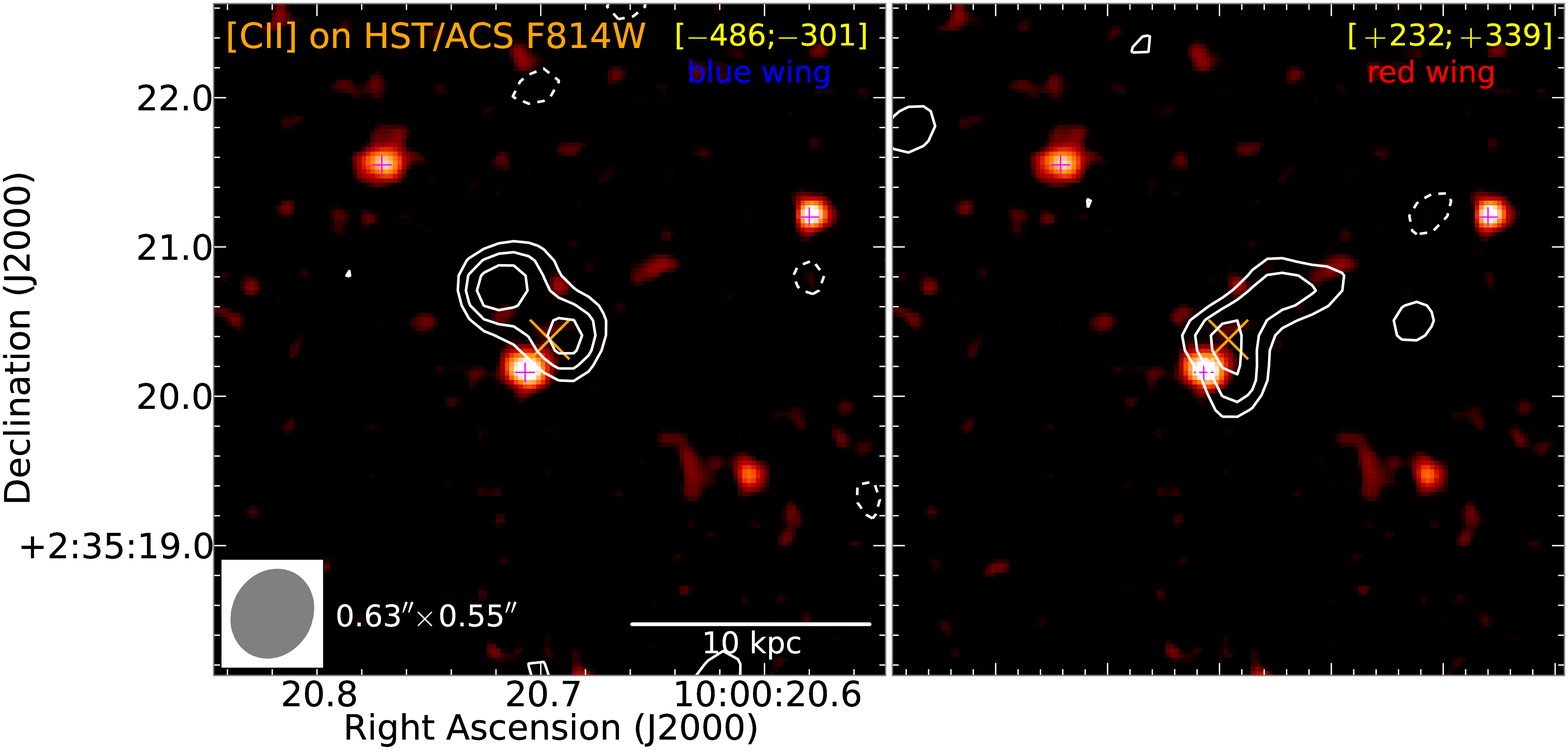}
\vspace{-4mm}

\caption{ALMA \cii\ velocity channel contours of the line wings
  overlaid on {\em HST}/ACS F814W image toward AzTEC-3.  Velocity
  channels in the {\em left} and {\em right} panels are averaged over
  $\sim$185.55 and 107.42\,MHz ($\sim$184 and 107\,\kms ),
  respectively. Contours start at $\pm$3$\sigma$ and are in steps of
  1$\sigma$ (1$\sigma$=316 and 415\,$\mu$Jy\,beam$^{-1}$ at the phase
  center).  Velocity ranges in \kms\ are indicated in the top right
  corner of each panel. The synthesized beam size is the same as in
  Fig.~\ref{f2}.  The velocity scale is the same as in
  Fig.~\ref{f2}. The crosses and plus signs indicate the same
  positions as in Fig.~\ref{f2}.  \label{f4a}}
%
\end{figure}

We have detected strong \cii\ line emission toward AzTEC-3
(Figure~\ref{f2}). From two-dimensional Gaussian fitting to the line
emission, we find a deconvolved [C{\scriptsize II}] source size of
0.63$''$$^{+0.09''}_{-0.09''}$$\times$0.34$''$$^{+0.10''}_{-0.15''}$
(3.9$\times$2.1\,kpc$^2$; Tab.~\ref{t2}). The [C{\scriptsize II}]
emission is compact over the entire velocity range, without evidence
for a significant shift in the centroid position between velocity
channels (Figures~\ref{f2}, \ref{f2a}, and \ref{f4}). The
[C{\scriptsize II}] peak position is consistent with those of the CO
and FIR continuum emission (R10), as well as with the peak of the
optical emission longward of observed-frame $\sim$2\,$\mu$m
(Fig.~\ref{f10}). There is tentative evidence in different velocity
intervals for emission that may extend beyond the central, compact
component that dominates the [C{\scriptsize II}] emission, but only at
low flux density levels. The most prominent of these features is a
faint [C{\scriptsize II}] emission component in the blue line wing
that extends north-east from the center of the galaxy (Fig.~\ref{f4a},
{\em left}). This component could correspond to a tidal feature, a
close galaxy companion in a minor merger event, or a [C{\scriptsize
    II}] outflow (or inflow). A similar, but less significant feature
is seen in the red [C{\scriptsize II}] line wing (Fig.~\ref{f4a}, {\em
  right}).  Observations at higher spatial resolution are required to
further resolve the detailed structure of the interstellar medium in
this massive starburst galaxy.

We have also detected \cii\ emission toward the triple Lyman-break
galaxy system LBG-1 (Figure~\ref{f3}). The emission is spatially
resolved on a scale of
1.21$''$$^{+0.41''}_{-0.69''}$$\times$0.95$''$$^{+0.68''}_{-0.37''}$
(7.5$\times$5.9\,kpc$^2$; Tab.~\ref{t2}), covering all three optical
emission regions (LBG-1a, LBG-1b, and LBG-1c; Fig.~\ref{f3}). The
[C{\scriptsize II}] emission thus is clearly resolved toward the three
components, which is seen more prominently in different velocity bins
along the line emission (Figures~\ref{f3} and \ref{f5}). No
\cii\ emission is detected toward LBG-2 or LBG-3.

The \cii\ line parameters and redshifts for all detected galaxies were
determined with four parameter Gaussian fits to the line and continuum
emission spectra (Table~\ref{t1}). The [C{\scriptsize II}] emission in
AzTEC-3 has a peak flux density of 18.4$\pm$0.5\,mJy at a line FWHM of
421$\pm$19\,\kms, corresponding to an integrated line flux of
8.21$\pm$0.29\,Jy\,\kms, and peaks at a redshift of
$z$=5.2988$\pm$0.0001 (Fig.~\ref{f1a}, {\em left}).  The
[C{\scriptsize II}] emission in LBG-1 has a peak flux density of
8.99$\pm$0.73\,mJy at a line FWHM of 218$\pm$24\,\kms, corresponding
to an integrated line flux of 2.08$\pm$0.18\,Jy\,\kms, and peaks at a
central redshift of $z$=5.2950$\pm$0.0002 (Fig.~\ref{f1b}, {\em
  left}). The central velocities and line shapes of the [C{\scriptsize
    II}] emission are different at the optical positions of the three
regions (Fig.~\ref{f1b}, {\em middle}).  The three components LBG-1a,
b, and c peak at velocities of $-$49$\pm$18, $-$3$\pm$12, and
$+$25$\pm$16\,\kms\ relative to the central redshift, with line FWHM
of 93$\pm$32, 152$\pm$29, and 250$\pm$41\,\kms, respectively
(Table~\ref{t1b}; Fig.~\ref{f1b}, {\em middle}). LBG-1a is the
faintest and narrowest component, and thus, does not show a clearly
spatially separated peak in the velocity-integrated [C{\scriptsize
    II}] map (Fig.~\ref{f3}). However, the [C{\scriptsize II}] line
emission clearly extends toward LBG-1a in maps of narrower velocity
bins (Fig.~\ref{f5}). No [C{\scriptsize II}] emission is detected
toward LBG-2 and LBG-3 down to 3$\sigma$ limits of 0.21 and
0.21\,Jy\,\kms, assuming the same line width and redshift as measured
for LBG-1.  We also stacked the spectra of LBG-2 and LBG-3, which did
not result in a detection either (Fig.~\ref{f1b}, {\em right}). We
caution that this stack only provides a $\sqrt{2}$ deeper
[C{\scriptsize II}] limit if these galaxies are at a common redshift.

\begin{deluxetable*}{lccccccc}
\tabletypesize{\scriptsize}
\tablecaption{Derived [C{\scriptsize II}] and continuum properties. \label{t2}}
\tablehead{
Target & $d$([C{\scriptsize II}]) & $d_{\rm phys}$([C{\scriptsize II}]) & $d$(FIR) & $d_{\rm phys}$(FIR) & SFR$_{\rm FIR}$\tablenotemark{a} & $\Sigma_{\rm SFR}$          & $M_{\rm dyn}^{\rm [C{\scriptsize II}]}$\tablenotemark{b} \\
       &                          & [kpc$\times$kpc] &          & [kpc$\times$kpc] & [\msol\,yr$^{-1}$] & [\msol\,yr$^{-1}$\,kpc$^{-2}$] & [10$^{10}$\,\msol ] }
\startdata
AzTEC-3     & 0.63$''$$^{+0.09''}_{-0.09''}$$\times$0.34$''$$^{+0.10''}_{-0.15''}$ & 3.9$\times$2.1    & 0.40$''$$^{+0.04''}_{-0.04''}$$\times$0.17$''$$^{+0.08''}_{-0.17''}$ & 2.5$\times$1.1 & 1100 & 530 & 9.7 \\
LBG-1       & 1.21$''$$^{+0.41''}_{-0.69''}$$\times$0.95$''$$^{+0.68''}_{-0.37''}$ & 7.5$\times$5.9    & {\em undetected} & --- & $<$18--54\tablenotemark{c} & --- & 5.0 \\
LBG-2       & {\em undetected}                                              & ---               & {\em undetected} & --- & $<$28\tablenotemark{d} & --- & --- \\
LBG-3       & {\em undetected}                                              & ---               & {\em undetected} & --- & $<$28\tablenotemark{d} & --- & --- \\
\vspace{-3mm}
\enddata
\tablenotetext{a}{Determined from $L_{\rm FIR}$, assuming a Chabrier (\citeyear{cha03}) initial mass function.}
\tablenotetext{b}{Derived assuming an isotropic virial estimator (e.g., Engel et al.\ \citeyear{eng10}), and the galaxy size along the major axis.}
\tablenotetext{c}{The lower of the limits assumes that all SFR$_{\rm FIR}$ is concentrated toward one component, the higher limit assumes that it is spatially resolved and spread over all components (LBG-1a, LBG-1b, and LBG-1c).}
\tablenotetext{d}{Point source limits, assuming the same redshift and SED shape as for LBG-1.}
\end{deluxetable*}


\subsection{OH and CO Line Emission}

\begin{figure*}
\epsscale{1.15}
\plotone{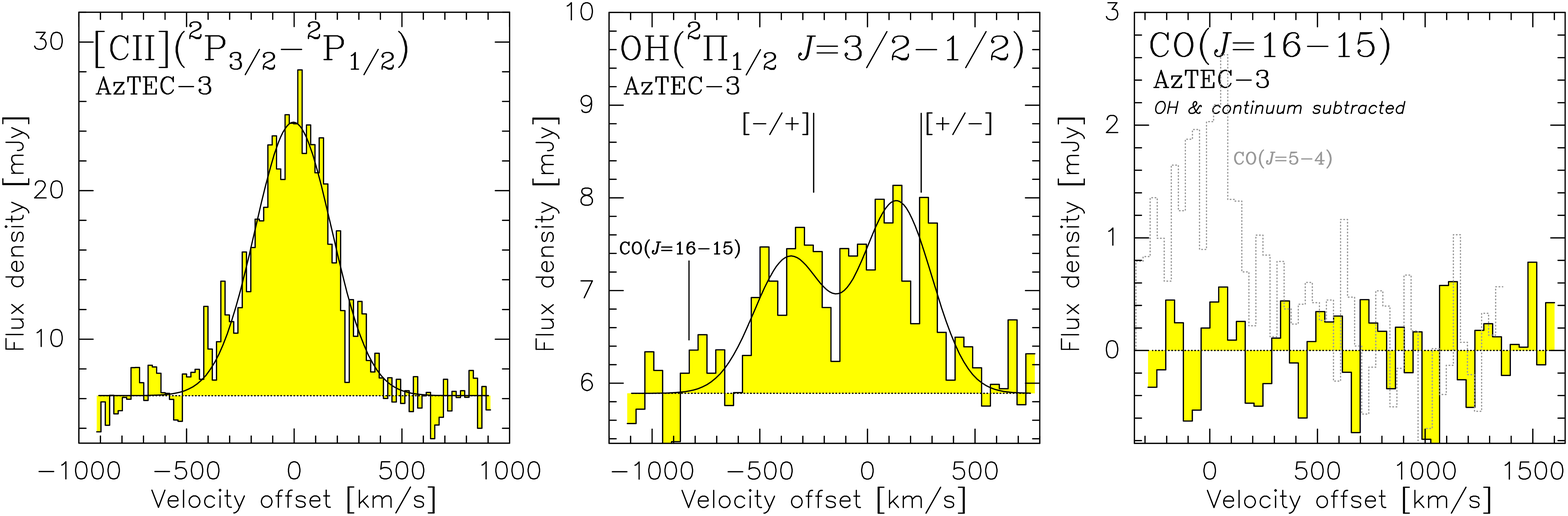}
\vspace{-1mm}

\caption{ALMA spectra of the \cii\ ({\em left}), \oh\ ({\em middle}),
  and \pco\ ({\em right}) lines toward AzTEC-3. Spectra (histograms)
  are shown at resolutions of 20\,MHz (20\,\kms; [CII]) or 40\,MHz
  (41\,\kms; OH and CO). The velocity scales are relative to
  $z$=5.2988. Detected lines are shown along with Gaussian fits to the
  line emission (black curves).  In the case of OH, zero velocity
  corresponds to the central frequency between the $P$ components of
  the $\Lambda$ doublet. The [+/--] and [--/+] labels at
  $\pm$249\,\kms\ indicate the $P$ components of the $\Lambda$ doublet
  (i.e., $J^P$=3/2$^+$$\to$1/2$^-$ and 3/2$^-$$\to$1/2$^+$,
  respectively). Due to the narrow splitting ($\sim$14 and 2\,MHz
  observed-frame, respectively), labels for hyperfine structure
  splitting of the $P$ components are omitted for clarity. OH and
  continuum emission as quantified by the Gaussian fit in the {\em
    middle} panel have been subtracted from the \pco\ spectrum. The
  dashed gray histogram shows the \eco\ emission (R10) for comparison,
  which demonstrates that the bulk of the \pco\ line is covered by the
  1.875\,GHz bandpass shown.
  \label{f1a}}
\vspace{-3mm}

\end{figure*}

We have successfully detected the \oh\ doublet toward AzTEC-3
(Fig.~\ref{f1a}, {\em middle}). We measure peak flux densities of
1.46$\pm$0.22 and 2.07$\pm$0.22\,mJy for the two components of the
$\Lambda$--doublet, at a common line FWHM of 384$\pm$43\,\kms\ for
each component. This corresponds to an integrated line flux of
1.44$\pm$0.13\,Jy\,\kms\ (Table~\ref{t1a}).  This 163\,$\mu$m OH
feature thus has a higher line flux than any CO lines detected in this
source, carrying almost 20\% of the flux of the \cii\ line. The
central velocity of the feature is shifted by
--109$\pm$19\,\kms\ relative to the [C{\scriptsize II}] line. This may
indicate that the OH emission is associated with an outflow, but
better sensitivity and higher spatial resolution observations are
required to further investigate such a scenario.

Our observations also covered the \pco\ line in AzTEC-3, but no
emission was detected (Fig.~\ref{f1a}, {\em right}). We place a
3$\sigma$ limit of $<$0.22\,Jy\,\kms\ on the strength of the line,
assuming the same width as that of the \eco\ line (487$\pm$58\,\kms;
R10). This limit is 4--6$\times$ lower than the fluxes measured in the
\eco\ and \fco\ lines, suggesting that no strong component with high
CO excitation is present.

Based on the [C{\scriptsize II}] redshift, we have used a recent data
set (PI:\ Riechers) from the Karl G.\ Jansky Very Large Array (VLA) to
search for \bco\ line emission toward LBG-1. We do not detect any
signal down to an approximate 3$\sigma$ limit of $<$0.03\,Jy\,\kms,
assuming the same width and redshift as for the [C{\scriptsize II}]
line (Fig.~\ref{f11}; these data will be described in detail in a
future publication; D.~A\ Riechers et al., in prep.).  Given this
non-detection, we do not consider the limits on the \oh\ and
\pco\ lines from the ALMA data to be constraining for this
source. Given the non-detections of [C{\scriptsize II}], we do not
extract limits on these lines for LBG-2 and LBG-3 either.


\begin{deluxetable}{ l c c }
\tabletypesize{\scriptsize}
\tablecaption{[C{\scriptsize II}] properties of LBG-1. \label{t1b}}
\tablehead{
Target & $v_0$\tablenotemark{a} & d$v_{\rm CII}$ \\
       & [\kms ] & [\kms ] }
\startdata
{\em LBG-1 (all components)} & ---          & {\em 218$\pm$24} \\
LBG-1a      & $-$49$\pm$18 &  93$\pm$32 \\
LBG-1b      & $-$3$\pm$12  & 152$\pm$29 \\
LBG-1c      & $+$25$\pm$16 & 250$\pm$41 \\
\vspace{-2mm}
\enddata 
\tablenotetext{a}{Central line velocity relative to a redshift of 5.2950.}

\end{deluxetable}



\begin{deluxetable}{ l c c c }
\vspace{-7mm}

\tabletypesize{\scriptsize}
\tablecaption{Line fluxes and luminosities 
in AzTEC-3. \label{t1a}}
\tablehead{
& $I_{\rm line}$ & $L'_{\rm line}$ & Ref. \\
& [Jy\,\kms ] & [10$^{10}$\,K\,\kms\,pc$^2$] & }
\startdata
\bco\   & 0.23 $\pm$ 0.03 & 5.84 $\pm$ 0.78 & 1 \\ 
\eco\   & 0.92 $\pm$ 0.09 & 3.70 $\pm$ 0.37 & 1 \\ 
\fco\   & 1.36 $\pm$ 0.19 & 3.82 $\pm$ 0.54 & 1 \\ 
\pco\   & $<$0.22\tablenotemark{a} & $<$0.09\tablenotemark{a} & 2 \\ 
\oh\    & 1.44 $\pm$ 0.13 & 0.57 $\pm$ 0.05 & 2 \\ 
\cii\   & 8.21 $\pm$ 0.29 & 3.05 $\pm$ 0.11 & 2 \\
\vspace{-2mm}
\enddata 
\tablecomments{All quoted upper limits are 3$\sigma$.}
\tablerefs{${}$[1] Riechers et al.\ (\citeyear{rie10}), [2] this work.
\tablenotetext{a}{Assuming the same width as measured for the \eco\ line (R10). 
Does not account for any uncertainty due to prior subtraction of OH emission 
(the closer OH component peaks $\sim$580\,\kms\ redwards of the line).}
}
\end{deluxetable}


\subsection{Spectral Energy Distribution}

\subsubsection{AzTEC-3}

To determine the spectral energy distribution (SED) properties of
AzTEC-3 after including the new constraints from ALMA, we have fit
modified black-body (MBB) models to the continuum data between
observed-frame 100\,$\mu$m and 8.2\,mm (R10; C11; Huang et
al.\ \citeyear{hua14}; Smol\v{c}i\'c et al.\ \citeyear{smo14}).  The
MBB is joined to a $\nu^{-\alpha}$ power law on the blue side of the
SED peak.

We used an affine-invariant Markov Chain Monte-Carlo (MCMC) approach,
employing the method described by Riechers et al.\ (\citeyear{rie13})
and Dowell et al.\ (\citeyear{dow14}).\footnote{\texttt
  https://github.com/aconley/mbb\_emcee} We first fit optically thin
models, using $\alpha$, the dust temperature $T_{\rm dust}$, the power
law slope of the dust extinction curve $\beta$, and a normalization
factor (for which we elect the observed-frame 500\,$\mu$m flux
density) as fitting parameters. A weak Gaussian prior of 1.9$\pm$0.3
is adopted for $\beta$. The best fit solution has a $\chi^2$ of 19.44
for 10 degrees of freedom. We find $T_{\rm
  dust}$=52.8$^{+5.0}_{-5.4}$\,K, $\beta$=1.83$\pm$0.22, and
$\alpha$=6.64$^{+2.56}_{-2.48}$, but we note that $\alpha$ is only
poorly constrained by the data. The relatively high $\chi^2$ reflects
the fact that the routine experiences difficulties with simultaneously
fitting the short- and long-wavelength data with the given set of
parameters.

To improve the match, we also fit optically thick models, introducing
the wavelength $\lambda_0$=c/$\nu_0$ where the optical depth
$\tau_\nu$=($\nu$/$\nu_0$)$^\beta$ reaches unity as an additional
fitting parameter. The best fit solution has a $\chi^2$ of 14.63 for 9
degrees of freedom, suggesting a significantly better fit than in the
optically-thin case. We find $\lambda_0$=177$^{+39}_{-38}$\,$\mu$m
(rest-frame wavelength). We also find $T_{\rm
  dust}$=88.4$^{+9.8}_{-9.6}$\,K, $\beta$=2.16$\pm$0.27, and
$\alpha$=6.17$^{+2.73}_{-2.60}$. We caution that the uncertainties on
the photometry close to the peak of the SED are $>$30\%--50\% (see
also discussion by Huang et al.\ \citeyear{hua14}; Smol\v{c}i\'c et
al.\ \citeyear{smo14}), allowing for a broad range in possible peak
wavelengths. The best fit solution yields a FIR luminosity of $L_{\rm
  FIR}$=(1.10$^{+0.22}_{-0.21}$)$\times$10$^{13}$\,\lsol.\footnote{$L_{\rm
    FIR}$ is determined over the rest-frame 42.5--122.5\,$\mu$m range
  throughout. Note that C11 determined $L_{\rm FIR}$ for AzTEC-3 over
  the rest-frame 60--120\,$\mu$m range.} Assuming standard relations
and a dust absorption coefficient of $\kappa_\nu$=2.64\,m$^2$kg$^{-1}$
at 125\,$\mu$m (e.g., Dunne et al.\ \citeyear{dun03}, their equation
1),\footnote{The uncertainty in $\kappa_\nu$ is at least
  $\sim$0.4\,dex.} we also find a dust mass of $M_{\rm
  dust}$=2.66$^{+0.74}_{-0.80}$$\times$10$^8$\,\msol.

\begin{figure*}
\epsscale{0.93}
\plotone{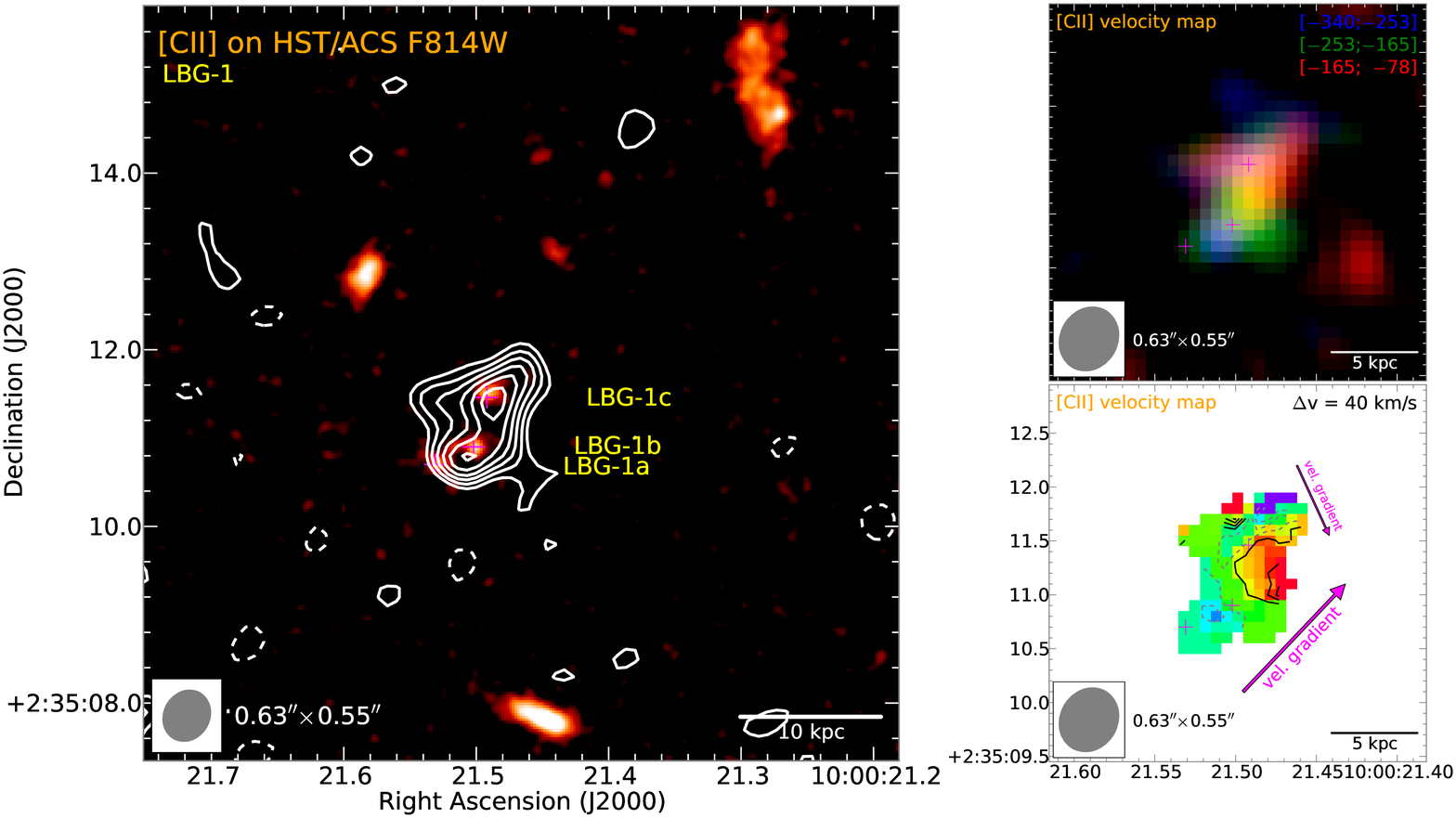}
\vspace{-2mm}

\caption{Velocity-integrated ALMA \cii\ contours overlaid on {\em
    HST}/ACS F814W image ({\em left}) and \cii\ velocity maps ({\em
    right}) toward LBG-1. {\em Left:}\ The map is averaged over
    $\sim$332\,MHz (330\,\kms ). Contours start at $\pm$3$\sigma$ and
    are in steps of 1$\sigma$ (1$\sigma$=236\,$\mu$Jy\,beam$^{-1}$ at
    the phase center). The synthesized beam size is the same as in
    Fig.~\ref{f2}. The plus signs indicate the same positions as in
    Fig.~\ref{f1}. {\em Right:}\ Color-encoded velocity structure and
    first moment map of the [C{\scriptsize II}] emission. The channel
    maps in the {\em top right} panel include emission above 3$\sigma$
    significance (1$\sigma$=460\,$\mu$Jy\,beam$^{-1}$ at the phase
    center). The colors indicate different velocity bins. Velocity
    ranges in \kms\ for each color are indicated in the top right
    corner. The velocity scale is the same as in Fig.~\ref{f2}. The
    colors in the first moment map ({\em bottom right}) indicate the
    velocity gradient. Contours are shown in steps of 40\,\kms, with
    dashed (solid) contours showing blueshifted (redshifted) emission
    relative to the central velocity of LBG-1 at $z$=5.2950. The
    arrows indicate the general directions (blue- to redshifted) of
    the two strongest velocity gradients.
\label{f3}}
%
\end{figure*}

\begin{figure*}
\epsscale{1.15}
\plotone{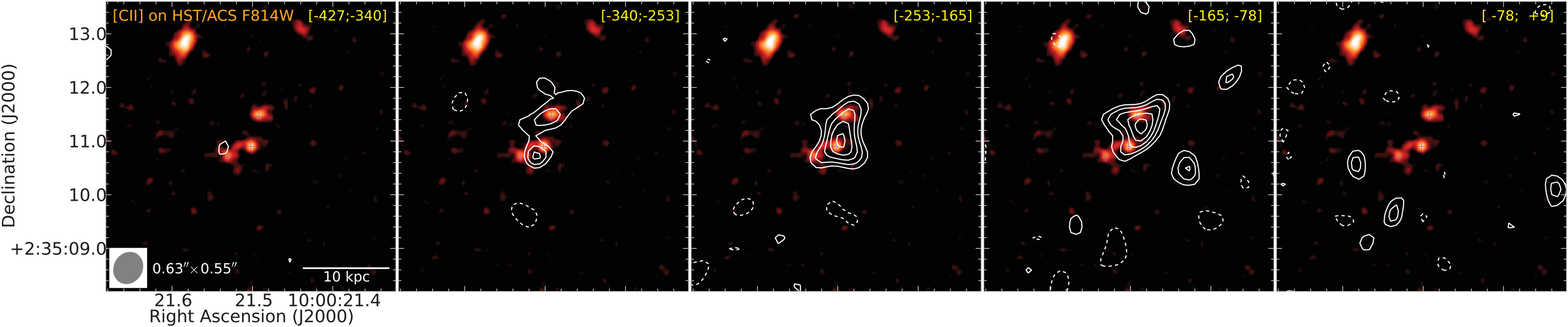}
\vspace{-2mm}

\caption{ALMA \cii\ velocity channel contours overlaid on {\em
    HST}/ACS F814W image toward LBG-1.  Velocity channels are averaged
  over 87.89\,MHz (87\,\kms ).  Contours start at $\pm$3$\sigma$ and
  are in steps of 1$\sigma$ (1$\sigma$=460\,$\mu$Jy\,beam$^{-1}$ at
  the phase center). The noise in the rightmost channel is slightly
  higher due to a weak atmospheric absorption feature (see Sect.~2).
  Velocity ranges in \kms\ are indicated in the top right corner of
  each panel. The synthesized beam size is the same as in
  Fig.~\ref{f3}.  The velocity scale is the same as in
  Fig.~\ref{f3}. The plus signs indicate the same positions as in
  Fig.~\ref{f3}.  \label{f5}}
%
\end{figure*}

\begin{figure*}
\epsscale{1.15}
\plotone{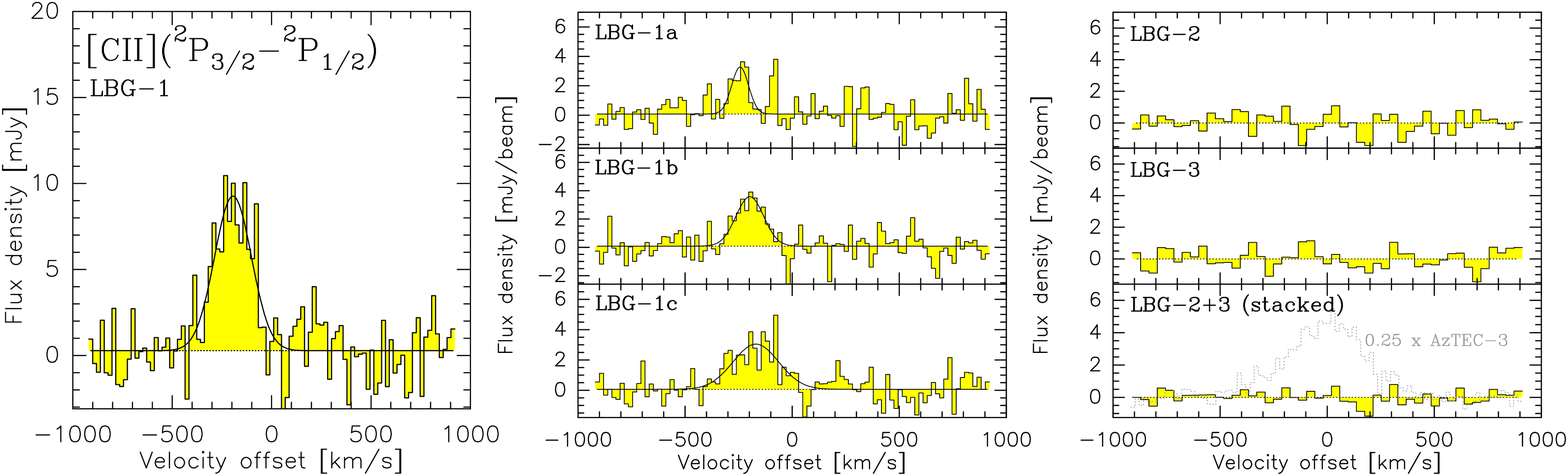}
\vspace{-1mm}

\caption{ALMA \cii\ spectra toward LBG-1 ({\em left}) and its
  components LBG-1a, LBG-1b, and LBG-1c ({\em middle}; see
  Fig.~\ref{f3}), as well as LBG-2 and LBG-3 ({\em right}). Spectra
  (histograms) are shown at resolutions of 19.5\,MHz (19\,\kms; {\em
    left} and {\em middle}) or 39\,MHz (39\,\kms; {\em right}). The
  {\em left} spectrum corresponds to the spatially-integrated emission
  shown as contours in Fig.~\ref{f3}. The {\em middle} and {\em right}
  spectra are extracted at the optical peak positions for all sources
  (plus signs in Figs.~\ref{f1} and \ref{f3}). The velocity scales are
  relative to $z$=5.2988. Detected lines are shown along with Gaussian
  fits to the line emission (black curves). The {\em bottom right}
  panel shows a stack of the spectra of LBG-2 and LBG-3. The dashed
  gray histogram shows the \cii\ emission from the strong nearby
  source AzTEC-3 (Fig.~\ref{f1a}), scaled by a factor of 0.25, for
  comparison.
  \label{f1b}}
%
\end{figure*}

\begin{figure}
\epsscale{1.15}
\plotone{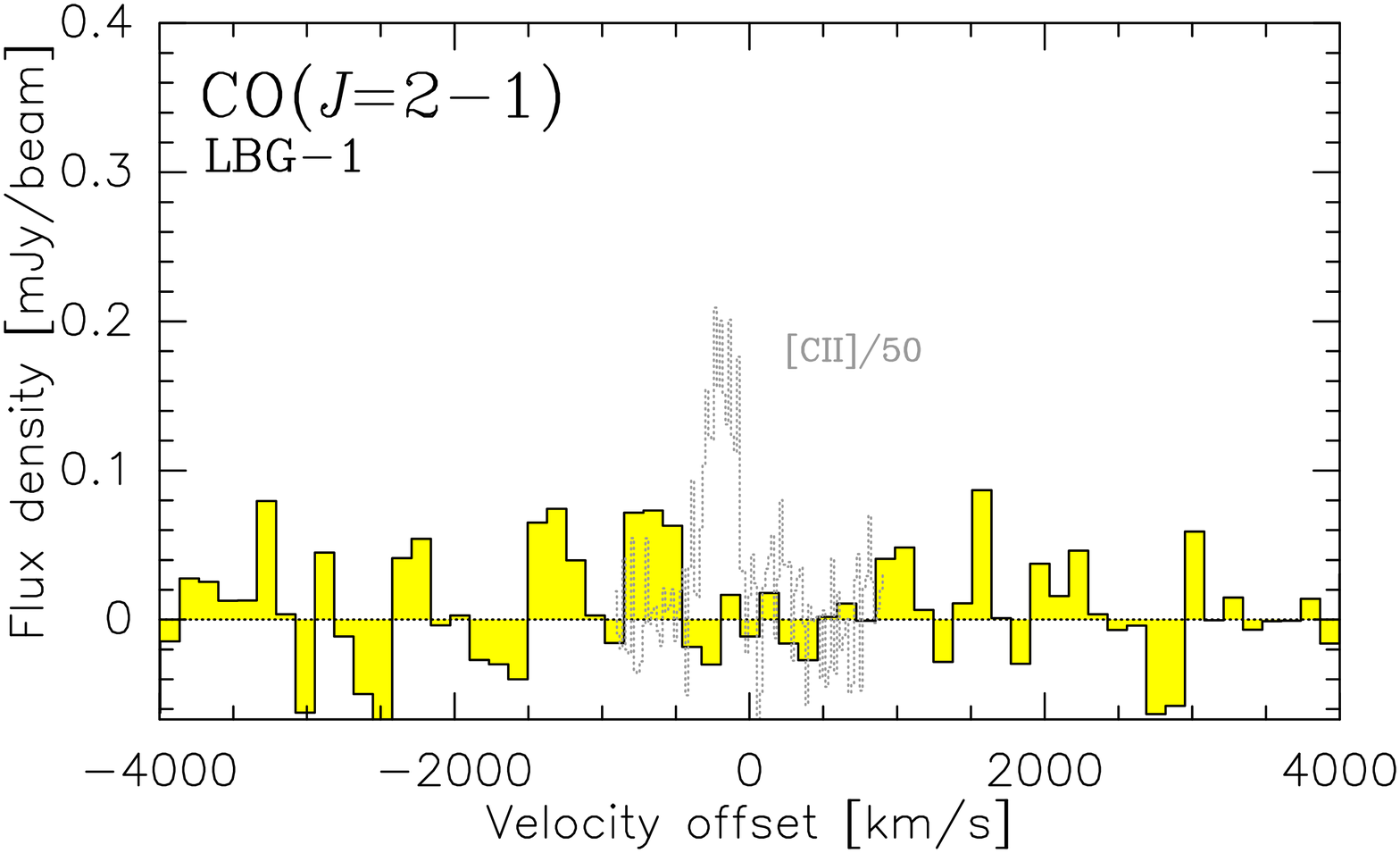}
\vspace{-1mm}

\caption{VLA spectrum covering the redshifted \bco\ frequency toward
  LBG-1 (D.A.~Riechers et al., in prep.). Spectrum (histogram) is
  shown at a resolution of 16\,MHz (131\,\kms ). The velocity scale is
  relative to $z$=5.2988. The dashed gray histogram shows the
  \cii\ emission (Fig.~\ref{f1b}), scaled by a factor of 1/50, for
  comparison. \label{f11}}
%
\end{figure}

\subsubsection{Lyman-Break Galaxies}

\begin{figure*}
\epsscale{1.15}
\plotone{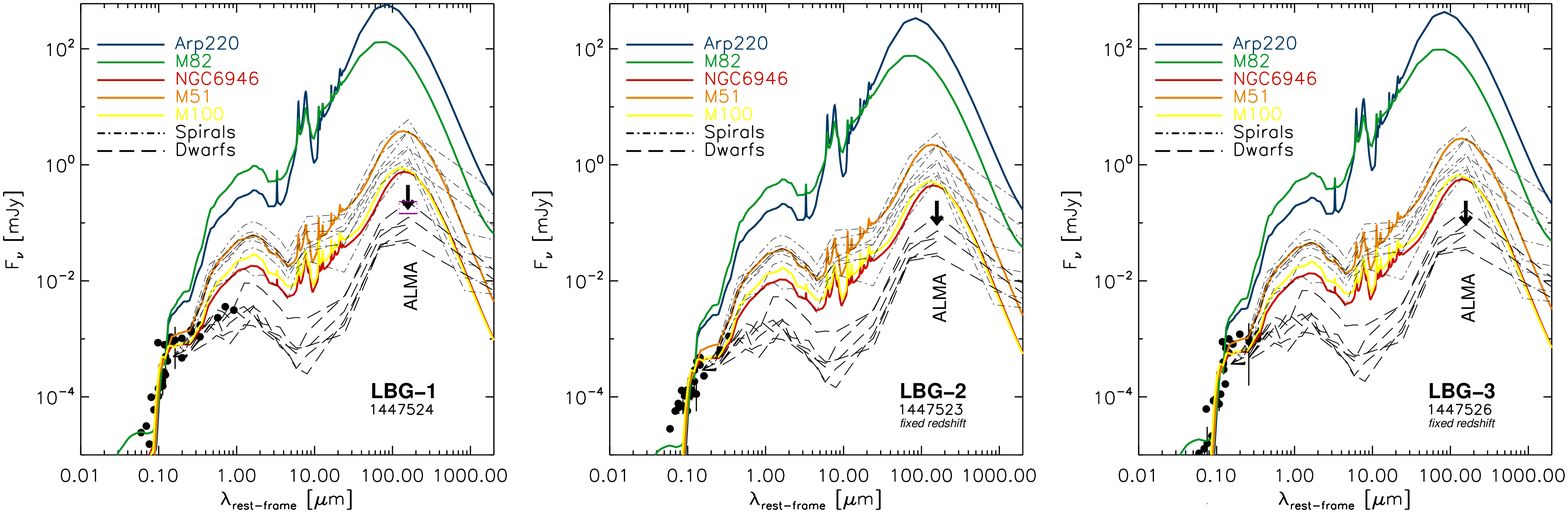}
\vspace{-2mm}

\caption{Spectral energy distributions of LBG-1, LBG-2, and LBG-3, and
  matched galaxy templates. The redshifts for LBG-2 and LBG-3 are
  assumed to be the same as measured for LBG-1. The templates for the
  nearby starburst and ultra-luminous infrared galaxies M51, M82,
  M100, NGC\,6946, and Arp\,220 (colored lines; Silva et
  al.\ \citeyear{sil98}), the spirals NGC\,1097, NGC\,3351, NGC\,5055,
  NGC\,3031, NGC\,7331, NGC\,4736, and NGC\,3627 (dot-dashed lines),
  and the dwarfs Holmberg\,I, NGC\,6822, Holmberg\,II, DDO\,053,
  IC\,2574, and NGC\,1705 (dashed lines; both samples from Dale et
  al.\ \citeyear{dal07}) are normalized to the Subaru $i$-band fluxes
  of the LBGs.  The arrows indicate the upper limits on the rest-frame
  157.7\,$\mu$m continuum flux obtained from our ALMA
  observations. The magenta bars in the {\em left} panel indicate the
  range of 157.7\,$\mu$m point source limits for the subcomponents
  LBG-1a, LBG-1b, and LBG-1c.  \label{f6}}
%
\end{figure*}

Given the detailed available photometry from COSMOS, we match galaxy
templates to the observed SED of LBG-1 (see Fig.~\ref{f6}, {\em
  left}). Fits are obtained by normalizing all templates to the
observed Subaru $i$-band flux of the LBG (i.e., rest-frame ultraviolet
light). Differences in photometry shortward of the Lyman--$\alpha$
line are not considered further in the comparison, since the nearby
galaxy templates do not account for intergalactic medium absorption at
$z$=5.3 due to the Gunn-Peterson effect (Gunn \& Peterson
\citeyear{gp65}). The ALMA continuum limit for LBG-1 is inconsistent
with the SED shapes of spiral, starburst and dust-obscured galaxies in
the nearby universe, but is consistent with the flatter SED shapes
typically observed in nearby dwarf galaxies. Similar results are found
for LBG-2 and LBG-3 when assuming the same redshift as for LBG-1
(Fig.~\ref{f6}, {\em middle} and {\em right}). To determine the
$L_{\rm FIR}$ of LBG-1, we thus integrated the dwarf galaxy templates
over their infrared peaks.\footnote{Corresponding to the rest-frame
  42.5--122.5\,$\mu$m range.} This results in $L_{\rm FIR}$ limits of
$<$1.1--3.4$\times$10$^{11}$\,\lsol.\footnote{We did not correct for
  effects due to the cosmic microwave background (CMB), which has a
  temperature of $T_{\rm z,CMB}$$\simeq$17\,K at the redshift of
  AzTEC-3 and LBG-1. Any corrections to the $L_{\rm FIR}$ limits
  required are small compared to other sources of uncertainty unless
  the dust temperature approaches $T_{\rm z,CMB}$ (see, e.g., da Cunha
  et al.\ \citeyear{dc13}). The dust temperatures of the dwarf
  galaxies used as templates are $T_{\rm dust}$$\simeq$26--37\,K
  (e.g., Israel et al.\ \citeyear{isr96b}; Skibba et
  al.\ \citeyear{ski11}), which would require corrections at the few
  per cent level at most.}  We conservatively scale the highest of
these templates to the ALMA limits in the following. We then assume
LBG-2 and LBG-3 to have the same redshift and SED shape as LBG-1 to
determine limits on their $L_{\rm FIR}$.

Based on the upper limit for the rest-frame 157.7\,$\mu$m continuum
flux density, we assume standard relations and the same dust
absorption coefficient as for AzTEC-3 to place constraints on the dust
mass of LBG-1. Assuming a dust temperature of $T_{\rm dust}$=30\,K and
an opacity power law index of $\beta$=1.5 yields $M_{\rm
  dust}$$<$3.1--9.4$\times$10$^7$\,\msol,\footnote{These assumptions
  for $T_{\rm dust}$ and $\beta$ are consistent with constraints
  obtained from stacking studies of LBGs at lower redshifts (e.g., Lee
  et al.\ \citeyear{lee12}). Assuming $T_{\rm dust}$=25 or 40\,K
  instead would yield $\sim$1.9$\times$ or $\sim$0.44$\times$ the
  quoted $M_{\rm dust}$ limit. Assuming $\beta$=2.0 instead would
  yield a 12\% higher limit for $M_{\rm dust}$, and thus, would have a
  minor impact compared to other sources of uncertainty.} where the
range indicates the difference between point source and extended
source limits.

\subsection{Derivation of Further Galaxy Properties}

We derive line luminosities from the observed [C{\scriptsize II}]
intensities using standard relations (e.g., Solomon \& vanden Bout
\citeyear{sv05}; Carilli \& Walter \citeyear{cw13}). 
We assume a Chabrier (\citeyear{cha03}) stellar initial mass function
(IMF) to derive star formation rates from $L_{\rm FIR}$.  For AzTEC-3,
we use the measured continuum size to estimate the average star
formation rate surface density. Based on the line width and
[C{\scriptsize II}] galaxy sizes measured along the major axis, we
then use an isotropic virial estimator (e.g., Engel et
al.\ \citeyear{eng10}) to derive dynamical masses for all
[C{\scriptsize II}]--detected sources (see Table~\ref{t2}).

\section{Analysis}

\subsection{The Massive Starburst Galaxy AzTEC-3}

Despite its high $L_{\rm FIR}$, both the continuum and [C{\scriptsize
    II}] line emission in the massive starburst galaxy AzTEC-3 are
fairly compact. The bulk of the atomic gas as traced by [C{\scriptsize
    II}] is distributed over a region of only $\lesssim$2\,kpc
radius.\footnote{This result is consistent with earlier estimates of
  the size of the molecular gas reservoir, which resulted in a limit
  of $<$4\,kpc on the radius based on lower spatial resolution CO
  observations. Under the plausible assumption that the system is not
  dark matter dominated within its central few kiloparsec, a limit on
  the gas disk-equivalent radius of $r_0$$<$2.3\,kpc was obtained
  (R10).} In spite of the high signal-to-noise ratio (and thus, high
centroid precision in the velocity channels in Fig.~\ref{f4}) of our
[C{\scriptsize II}] detection, there is no evidence for a significant
velocity gradient on $>$1\,kpc scales for the bulk of the
emission. This suggests that the comparatively large velocity widths
of the atomic and molecular lines are dominantly supported by emission
from highly-dispersed gas (the median [C{\scriptsize II}] velocity
dispersion at the spatial resolution of our observations is
$\sim$315\,\kms ), rather than ordered rotation. Only at faint levels,
we find evidence for tidal structure or outflowing/infalling streams
of gas.  AzTEC-3 thus appears to be considerably more compact than
some other $z$$>$4 SMGs such as GN20 or HDF\,850.1 ($\sim$4 times
larger in the FIR continuum; Younger et al.\ \citeyear{you08}; Carilli
et al.\ \citeyear{car10}; Walter et al.\ \citeyear{wal12b}; Neri et
al.\ \citeyear{ner14}), but it is more comparable in extent to the
$z$=6.34 starburst HFLS3 (3.4\,kpc$\times$2.9\,kpc diameter in
[C{\scriptsize II}], 2.6\,kpc$\times$2.4\,kpc diameter in the
rest-frame 157.7\,$\mu$m continuum; Riechers et
al.\ \citeyear{rie13}).

Based on the size of the far-infrared continuum emission
($\sim$2.1\,kpc$^2$) and the star formation rate of
1100\,\msol\,yr$^{-1}$, we find a high star formation rate surface
density of $\Sigma_{\rm SFR}$=530\,\msol\,yr$^{-1}$\,kpc$^{-2}$ and a
$L_{\rm FIR}$ surface density of $\Sigma_{\rm
  FIR}$=0.5$\times$10$^{13}$\,\lsol\,kpc$^{-2}$. This is close to the
theoretically predicted Eddington limit for starburst disks that are
supported by radiation pressure, and is consistent with theories for
so-called ``maximum starbursts'' (Elmegreen \citeyear{elm99}; Scoville
\citeyear{sco03}; Thompson et al.\ \citeyear{tho05}). Comparable,
compact ``hyper-starbursts'' were found in the $z$=6.42 quasar host
galaxy J1148+5251, and in the $z$=6.34 dusty galaxy HFLS3
($\Sigma_{\rm SFR}$=1000 and 600\,\msol\,yr$^{-1}$\,kpc$^{-2}$; Walter
et al.\ \citeyear{wal09}; Riechers et al.\ \citeyear{rie13}). Nearby,
comparable $\Sigma_{\rm SFR}$ are only found in the very centers of
GMCs or in the nuclei of ultra-luminous infrared galaxies like
Arp\,220.

The picture of a relatively compact, warm starburst is consistent with
the molecular gas excitation in AzTEC-3, which is comparable to that
in the similarly compact and warm $z$=6.34 starburst HFLS3 (Riechers
et al.\ \citeyear{rie13}), but considerably higher than in other, less
compact $z$$>$4 SMGs like GN20 and HDF\,850.1 (R10; Carilli et
al.\ \citeyear{car10}; Walter et al.\ \citeyear{wal12b}) and in
typical $z$$\sim$2--3 SMGs (e.g., Riechers et al.\ \citeyear{rie11a},
\citeyear{rie11b}). The gas excitation in AzTEC-3, however, is lower
than in quasar host galaxies, where the active galactic nucleus (AGN)
may contribute to the excitation of high-level CO lines (e.g.,
Riechers et al.\ \citeyear{rie06}, \citeyear{rie09b},
\citeyear{rie11c}; Wei\ss\ et al.\ \citeyear{wei07}). In particular,
the non-detection of \pco\ emission is consistent with no luminous AGN
component contributing to the gas heating, in agreement with previous
models of the CO excitation (R10). This picture is also consistent
with the relatively low $L_{\rm CII}$/$L_{\rm FIR}$ ratio of
4.0$\times$10$^{-4}$. This value is comparable to that of HFLS3
(5.4$\times$10$^{-4}$; Riechers et al.\ \citeyear{rie13}), and lower
than in other, less compact $z$$>$4 SMGs like HDF\,850.1
(1.7$\times$10$^{-3}$; Walter et al.\ \citeyear{wal12b}), but higher
than in some quasar host galaxies (e.g., 1.9$\times$10$^{-4}$ for
J1148+5251; Walter et al.\ \citeyear{wal09}). It is in agreement with
the ($L_{\rm CII}$/$L_{\rm FIR}$)--$\Sigma_{\rm IR}$ relation for
nearby infrared-luminous galaxies (Diaz-Santos et
al.\ \citeyear{ds13}). This ratio suggests the presence of a
stronger-than-average far-UV radiation field due to a warm, dense,
compact starburst with high $\Sigma_{\rm SFR}$, but gives no direct
indication for the presence of an obscured AGN. Given the inferred
compactness and high density of the gas and dust, it cannot be ruled
out that the optical depth of the dust in AzTEC-3 at rest-frame
157.7\,$\mu$m is considerable, and thus, that $L_{\rm CII}$ is
comparatively low due to extinction. Using the $L_{\rm CO(1-0)}$ found
from CO excitation modeling to \bco\ and higher-$J$ lines (R10), we
find $L_{\rm CII}$/$L_{\rm CO(1-0)}$$\simeq$2070, which is by
$\sim$50\% lower than the value found for HFLS3 ($L_{\rm CII}$/$L_{\rm
  CO(1-0)}$$\simeq$3050; Riechers et al.\ \citeyear{rie13}). This
would be consistent with a higher dust optical depth in AzTEC-3
compared to HFLS3, resulting in a reduced [C{\scriptsize II}] line
luminosity. From our SED modeling, we find a dust optical depth of
$\tau_{\rm 157.7\,{\mu}m}$=1.28$\pm$0.04 at the wavelength of the
[C{\scriptsize II}] emission for AzTEC-3, which is higher than what is
found for HFLS3 ($\tau_{\rm 157.7\,{\mu}m}$$\lesssim$1; Riechers et
al.\ \citeyear{rie13}). This suggests that, in contrast to HFLS3, the
dust in AzTEC-3 is at least moderately optically thick towards
[C{\scriptsize II}] line emission. Extinction of the [C{\scriptsize
    II}] line by dust would be consistent with the finding that this
line may be somewhat narrower than the CO lines (421$\pm$19
vs.\ 487$\pm$58\,\kms; R10). However, it is important to point out
that internal, differential gas excitation between different
components within the galaxy could easily account for such a small,
barely significant difference in line width as well. We do not correct
the [C{\scriptsize II}] line luminosity for extinction, but we note
that the above picture would remain essentially unchanged for even a
factor of a few higher $L_{\rm CII}$.

The observed $L'_{\rm line}$ ratio, and thus, the brightness
temperature ($T_{\rm b}$) ratio between the \cii\ and \bco\ lines is
0.52$\pm$0.07 (Table \ref{t1a}). Previous two-component CO excitation
modeling suggests optical depths of $\tau_{\rm CO\,2-1}$=1.5 and 35.9
for the low- and high-excitation gas components that are estimated to
contribute 27.5\% and 72.5\% to the \bco\ line luminosity,
respectively (obtained from the models shown by R10), which implies
that the CO emission is optically thick. Assuming that the
[C{\scriptsize II}] and CO emission emerge from regions of similar
surface area, the similarity in observed $T_{\rm b}$ would suggest
that the [C{\scriptsize II}] line emission either is optically thick
as well, or alternatively, that it has an intrinsically higher line
excitation temperature than the CO emission ($T_{\rm ex}^{\rm
  CO\,2-1}$=18.3 and 44.9\,K at kinetic temperatures of $T_{\rm
  kin}$=30 and 45\,K for the low- and high-excitation gas components
in the model of R10). The latter may be expected if a large fraction
of the [C{\scriptsize II}] emission is associated with PDRs (e.g.,
Stacey et al.\ \citeyear{sta10}).

Finally, the detection of strong \oh\ emission at almost 20\% of the
[C{\scriptsize II}] line luminosity is also consistent with a warm
starburst with an intense radiation field. The high upper level energy
and high critical density of this transition ($E_{\rm up}$/$k_{\rm
  B}$$\simeq$270\,K and $n_{\rm crit}$$\simeq$10$^{8.6}$\,cm$^{-3}$,
respectively) makes collisional excitation unlikely, but instead
indicates the presence of a strong infrared radiation field. This is
consistent with what was found for the $z$=6.34 dusty starburst HFLS3,
the only other high-$z$ galaxy detected in the 163\,$\mu$m OH doublet
to date (Riechers et al.\ \citeyear{rie13}). The slight blueshift of
the OH feature relative to the CO and [C{\scriptsize II}] lines in
AzTEC-3 would be consistent with the presence of a molecular outflow.

We find a molecular gas mass fraction of $f_{\rm gas}$=$M_{\rm
  H_2}$/$M_{\rm dyn}$$\simeq$55\%, with $\sim$10\% of $M_{\rm dyn}$
being due to stellar mass. Taken at face value, this would indicate
that $\lesssim$35\% of the mass of AzTEC-3 in the central $\sim$4\,kpc
are due to dark matter. Conversely, even if no dark matter were to be
present, these considerations place an upper limit on the CO
luminosity to gas mass conversion factor of $\alpha_{\rm
  CO}$$<$1.3\,\msol\,(K\,\kms\,pc$^2$)$^{-1}$, which is
$\sim$3$\times$ lower than the Galactic $\alpha_{\rm
  CO}$.\footnote{The original molecular gas mass estimate of
  5.3$\times$10$^{10}$\,\msol\ assumed $\alpha_{\rm
    CO}$$=$0.8\,\msol\,(K\,\kms\,pc$^2$)$^{-1}$ for AzTEC-3 (R10).}
Using $\alpha_{\rm CO}$$=$0.8
($<$1.3)\,\msol\,(K\,\kms\,pc$^2$)$^{-1}$, we find a gas-to-dust mass
ratio of $M_{\rm H_2}$/$M_{\rm dust}$$\simeq$200 ($<$330), i.e.,
$\sim$3$\times$ higher than in HFLS3 when assuming the same
$\alpha_{\rm CO}$ (Riechers et al.\ \citeyear{rie13}). These values
are higher than the average of 120$\pm$28 found for the nuclei of
nearby infrared-luminous galaxies, but within the range of values
measured for individual sources (29--725; Wilson et
al.\ \citeyear{wil08}).

AzTEC-3, as the most massive, most intensely star-forming galaxy in
the protocluster, thus is a compact, gas-dominated galaxy that hosts a
maximal starburst. Its relative compactness, and implied high
$\Sigma_{\rm SFR}$ appear not to be a common feature among $z$$>$4
SMGs, but are comparable to the most extreme other cases known. In
particular, the $z$=4.05 galaxy GN20, which also is associated with an
overdensity of galaxies (Daddi et al.\ \citeyear{dad09}), is
significantly more extended by a factor of a few and has lower
$\Sigma_{\rm SFR}$ and lower molecular gas excitation (e.g., Carilli
et al.\ \citeyear{car10}). The $z$=5.18 SMG HDF\,850.1 is also
associated with a (less pronounced) overdensity of galaxies, is
significantly more extended by a factor of a few, and has lower
$\Sigma_{\rm SFR}$ and lower molecular gas excitation than AzTEC-3
(Walter et al.\ \citeyear{wal12b}). Assuming that these systems are
the progenitors of the same population of massive central cluster
galaxies, AzTEC-3 thus appears to be in a different phase of its
evolution. This makes it particularly interesting to further
understand how this may be connected to the properties of its
environment.

\subsection{The Lyman-Break Galaxies (LBGs)}

We solidly detect and spatially resolve [C{\scriptsize II}] emission
in the ``typical'' $z$$>$5 star-forming system LBG-1. The atomic gas
is spatially and dynamically resolved toward all three optically
identified sources. We resolve at least two major velocity gradients
between the different components (Figure~\ref{f3}). The smaller
gradient appears to be associated with the northern component
LBG-1c. The larger gradient appears to extend from the southern
component LBG-1a across the middle component LBG-1b toward LBG-1c
(blue- to redshifted emission), but it may mask smaller-scale
gradients associated with the individual components at the present
spatial resolution. In particular, there is tentative evidence for
smaller-scale velocity gradients with opposing directions, pointing
from the ``overlap'' region in between LBG-1a and LBG-1b towards the
centers of the two optical sources, but higher spatial resolution is
required to confirm and more clearly separate these smaller-scale
gradients.\footnote{Note that the spatial separation of the optical
  components in LBG-1 is not much larger than the synthesized beam
  size of our observations, but the centroid positions in individual
  velocity channel maps are known to significantly higher precision
  than the beam size at the given signal-to-noise ratio. This enables
  studies of velocity gradients on smaller scales than the beam size.}
This system thus may represent a merger of three Lyman-break galaxies,
or some other form of complex, clumpy, extended system in formation.
The dynamical mass of the triple system is about half that of the SMG
AzTEC-3 (Table~\ref{t2}), showing that LBG-1 contributes a significant
fraction of the mass bound in galaxies to the protocluster
environment.

The [C{\scriptsize II}] emission integrated over all components is
$\sim$4$\times$ fainter than in the SMG AzTEC-3, suggesting that each
of the components is typically by about an order of magnitude fainter
than the SMG. There is no evidence for continuum emission at the
position of LBG-1, suggesting that it is at least $\gtrsim$15$\times$
fainter ($>$25--40$\times$ fainter for the individual subcomponents)
than the SMG (3$\sigma$). Assuming an SED shape similar to AzTEC-3,
this would suggest a SFR of SFR$_{\rm FIR}$$<$80\,\msol\,yr$^{-1}$
($\lesssim$25--40\,\msol\,yr$^{-1}$ for the subcomponents). From the
[C{\scriptsize II}] luminosity, we obtain an estimate of SFR$_{\rm
  FIR}$$\sim$150\,\msol\,yr$^{-1}$, using the relation for dusty
starbursts recently suggested by Sargsyan et al.\ (\citeyear{sar12}).
This is somewhat inconsistent with the continuum-based estimate,
unless a bottom-heavy IMF is assumed (e.g., Conroy \& van Dokkum
\citeyear{cvd12}). However, despite the limited constraints on the
rest-frame far-infrared SED of the source, typical spiral/starburst
and dusty galaxy SED templates can be ruled out to fit the SED of
LBG-1 (Fig.~\ref{f6}). Templates for dwarf galaxies provide a
substantially better match to the overall SED, suggesting that LBG-1
is not a (very) dusty system, and providing an upper limit to the SFR
of SFR$_{\rm FIR}$$<$18--54\,\msol\,yr$^{-1}$ (the range represents
the difference between point source and extended source limits, since
the actual limit depends on the size of the emitting region; see
Table~\ref{t2}). We consider this to be the currently best estimate
for the FIR SFR in this system. From the range of templates that best
represent the data, the possibility of a comparatively low, sub-solar
metallicity cannot be excluded as an important reason for the low
implied FIR flux of the system. However, the detection of strong
157.7\,$\mu$m [C{\scriptsize II}] emission and the presence of deep
interstellar metal absorption features in rest-frame UV spectra (C11)
is inconsistent with very low metallicities. We thus consider it most
likely that the SED shape is due to the combination of a young
starburst and some contribution of optical light from a stellar
population that is already in place.  In particular, the lack of a
strong ``bump'' at rest-frame optical wavelengths (relative to the
strength of the rest-frame UV emission that is dominated by young,
massive stars) may indicate the lack of a significant older stellar
population. Such a scenario is consistent with what is expected for a
young, recently formed galaxy observed at an early cosmic epoch, only
1.1 billion years after the Big Bang. We independently estimated the
star formation rate based on the rest-frame UV continuum emission of
the galaxy (e.g., Madau et al.\ \citeyear{mad98}), using the SED
fitting methods outlined by C11.  This yields SFR$_{\rm
  UV}$=22\,\msol\,yr$^{-1}$. This is comparable to the best SFR$_{\rm
  FIR}$ limit, and again consistent with low dust extinction. By
assuming SFR$_{\rm total}$=SFR$_{\rm UV}$+SFR$_{\rm FIR}$, we find
that at most $<$45\%--71\% of the star formation activity in LBG-1 is
obscured by dust. The UV-to-optical SED shape is consistent with a
dust extinction of only $A_{\rm V}$$\simeq$0.5\,mag (C11). The finding
of low dust extinction is consistent with recent estimates for LBGs at
comparable and higher redshifts in the rest--frame optical (e.g.,
Bouwens et al.\ \citeyear{bou10}), suggesting that detection of such
galaxies in the FIR continuum may require substantial integration
times even with ALMA. This is also consistent with recent, less
sensitive far-infrared studies of Lyman-$\alpha$ emitters (Walter et
al.\ \citeyear{wal12}; Gonzalez-Lopez et al.\ \citeyear{gl14}). Given
its stellar mass (C11) and total star formation rate, LBG-1 is
consistent with being situated on the star-forming ``main sequence''
at $z$$\approx$5 (e.g., Speagle et al.\ \citeyear{spe14}).

From the [C{\scriptsize II}] luminosity and $L_{\rm FIR}$ limit, we
find a $L_{\rm CII}$/$L_{\rm FIR}$ limit of $>$3.1$\times$10$^{-3}$ to
$>$9.4$\times$10$^{-3}$ (the higher limit applies if the far-infrared
continuum emission is not resolved by our observations), i.e.,
approximately $>$0.3\% to $>$0.9\%.  This is consistent with normal,
star-forming galaxies at low redshift, but significantly higher than
in dusty quasars and extreme starburst galaxies (e.g., Gracia-Carpio
et al.\ \citeyear{gra11}; Diaz-Santos et al.\ \citeyear{ds13}).
Assuming thermalized excitation between \bco\ and \aco, we find a
$L_{\rm CII}$/$L_{\rm CO(1-0)}$ limit of $>$4600. This is consistent
with a comparatively modest UV radiation field strength (e.g., Stacey
et al.\ \citeyear{sta10}), comparable to normal, star-forming galaxies
in the local universe when accounting for the reduction in
\bco\ luminosity due to the CMB at the redshift of LBG-1.\footnote{We
  here assume a moderate gas kinetic temperature of $T_{\rm
    kin}$$\sim$30\,K, and that the \bco\ emission is optically thick.}

Our findings are consistent with the assumption that the protocluster
member LBG-1 is a ``typical'', close to $L^\star_{\rm UV}$
star-forming system at $z$$>$5 with relatively low dust content. LBG-1
appears to be a triple system, but our sensitivity would have been
sufficient to pick up a single component with one-third of the
[C{\scriptsize II}] luminosity within the protocluster. We thus
conclude that LBG-2 and LBG-3, which have photometric redshifts/colors
consistent with the protocluster environment, either have
[C{\scriptsize II}] luminosities that are lower than those of the
individual components of LBG-1 by at least a factor of 3 (which would
be consistent with their UV/optical properties within the relative
uncertainties), or have redshifts outside of our bandpass
(d$z$$\simeq$0.04 per 1.875\,GHz band). Independent spectroscopic
confirmation and/or more sensitive [C{\scriptsize II}] observations
will be required to distinguish between these possibilities. Note that
8 additional LBGs identified as likely protocluster members at larger
distances from the center (C11) were not covered by this initial
investigation, and thus, could have higher [C{\scriptsize II}] and/or
FIR luminosities than LBG-1.

\section{Discussion and Conclusions}

We have detected [C{\scriptsize II}] emission toward the intensely
star-forming SMG (AzTEC-3) and a triple Lyman-break galaxy system
(LBG-1) associated with the $z$=5.3 AzTEC-3 protocluster environment
(C11, R10). These member galaxies lie within a redshift range of
d$z$$<$0.004, suggesting that the association of galaxies is not only
close in the sky plane, but along the line of sight as well.  We
further detected \oh\ line and rest-frame 157.7\,$\mu$m continuum
emission in the SMG AzTEC-3, and placed a stringent limit on its
\pco\ luminosity. Our observations are consistent with a relatively
compact ($\sim$2.5\,kpc diameter), highly-dispersed, warm, ``maximum
starburst'' in its peak phase in the massive galaxy AzTEC-3, with
possible evidence for outflowing gas from the star-forming regions,
and/or tidal structure.  There is no evidence for an AGN contribution
to the excitation of the gas.  Its overall properties are reminiscent
of the most extremely active massive, dusty starburst galaxies found
within the (general) SMG population (e.g., Riechers et
al.\ \citeyear{rie13}).  Our observations are also consistent with
LBG-1 being a ``typical'', close to $L^\star_{\rm UV}$ galaxy
following the star-forming ``main sequence'' at its cosmic epoch, with
little evidence for old stellar populations or the presence of
dust. LBG-1 is not detected in sensitive CO or far-infrared continuum
observations, which is consistent with what is expected for a young
starburst with perhaps sub-solar metallicity. LBG-1 shows a complex
kinematic structure, perhaps representing a merger of three smaller
galaxies. Using only a fraction of the full ALMA science array, we
thus detect and spatially resolve the interstellar medium in both
distant massive starbursts and ``typical'' $z$$>$5 star-forming
galaxies at relative ease. We, however, do not detect two fainter
candidate members of the protocluster, which suggests that they either
are an order of magnitude fainter in [C{\scriptsize II}] emission than
LBG-1, or that their systemic redshifts fall outside the range covered
by our observations. The capabilities of ALMA in a more advanced stage
of completion will be necessary to further address this issue.

The detection of [C{\scriptsize II}] emission in LBG-1 is interesting
for a number of reasons. Recent searches for [C{\scriptsize II}]
emission in $z$$>$6.5 Lyman-$\alpha$ emitters (LAEs) and
Lyman-$\alpha$ blobs (LABs) have been unsuccessful, even with ALMA
(e.g., Walter et al.\ \citeyear{wal12}; Ouchi et
al.\ \citeyear{ouc13}; Gonzalez-Lopez et al.\ \citeyear{gl14}). These
systems have SFR$_{\rm UV}$ that are comparable to LBG-1, and even
exceed it by a factor of a few in the most extreme cases, but they
remain undetected in [C{\scriptsize II}] emission down to comparable,
and sometimes deeper levels than required to detect LBG-1. The
successful detection of LBG-1 may suggest that this is a selection
effect. The difference in cosmic time between the redshift of LBG-1
and $z$=6.5 is only $\sim$250\,Myr. As such, it is not clear that the
earlier epochs in which the LAEs and LABs were observed are the main
deciding factor. We consider it more likely that the narrow-band
Lyman-$\alpha$ selection technique that has led to the initial
identification of the LAEs and LABs targeted in [C{\scriptsize II}]
emission at $z$$>$6.5 selects against galaxies with sufficient
metallicity, and thus carbon abundance, to produce enough
[C{\scriptsize II}] line flux to be detectable. This is consistent
with the detection of [C{\scriptsize II}] emission in other, albeit
significantly more active and massive star-forming galaxy populations
at comparable redshifts (e.g., Riechers et al.\ \citeyear{rie13}).
Gonzalez-Lopez et al.\ (\citeyear{gl14}) have suggested that
high-redshift galaxies with high UV continuum fluxes but low
Lyman-$\alpha$ equivalent widths may be more likely to have sufficient
metallicity to be detectable in the [C{\scriptsize II}] line. This is
consistent with the [C{\scriptsize II}] detection of LBG-1, which has
a comparatively low Lyman-$\alpha$ equivalent width (C11). In any
case, our study shows that, even if the detection of dust in
``typical'' galaxies at very high redshift may require substantial
integration times, their gas content appears to be detectable with
ALMA in the [C{\scriptsize II}] line using only moderate amounts of
observing time, with the possible exception of systems with the lowest
metallicities (cf.\ Fisher et al.\ \citeyear{fis14}).

It is also interesting to discuss the properties of LBG-1 in the
context of the BR\,1202--0725 system at $z$=4.69. This system consists
of two far-infrared-luminous (each $>$10$^{13}$\,\lsol ), massive
galaxies, separated by 26\,kpc in projection, one of which is an
optically-luminous broad absorption line quasar. Early ALMA
observations have revealed two faint [C{\scriptsize II}]-emitting
sources in close proximity to the quasar ($<$15\,kpc in projection),
which appear to be associated with LAEs (components Ly$\alpha$-1 and
Ly$\alpha$-2; e.g., Wagg et al.\ \citeyear{wag12}; Carilli et
al.\ \citeyear{car13}). These LAEs, however, have very large
Lyman-$\alpha$ equivalent widths due to Lyman-$\alpha$ lines with
$>$1200\,\kms\ FWHM (e.g., Williams et al.\ \citeyear{wil14}).  Given
the strong tidal forces, and perhaps ongoing interaction between the
two massive galaxies, the close proximity and characteristics of these
LAEs thus have likely implications for the evolution (e.g., besides
its strong radiation field, the quasar shows a possible outflow in the
direction of one of the LAEs), and perhaps even the origin of these
sources (e.g., it cannot be ruled out that they represent, or formed
out of, tidal debris from the massive galaxies). As such, it is
unclear to what degree the LAEs in this system can be considered
``typical'' galaxies. However, given the lack of detections of
``typical'' high-redshift galaxies in [C{\scriptsize II}] emission
prior to our study, and since these LAEs are significantly less
extreme than all other systems detected in [C{\scriptsize II}] at high
redshift, it is instructive to compare the properties of LBG-1 to
those of the LAEs in BR\,1202--0725 (we adopt their properties from
the study of Carilli et al.\ \citeyear{car13} below).

In contrast to LBG-1, the Ly$\alpha$-2 component in the BR\,1202--0725
system is detected in the far-infrared continuum, suggesting a FIR
luminosity in excess of 10$^{12}$\,\lsol. Only part of the
[C{\scriptsize II}] line in Ly$\alpha$-2 is detected at the edge of
the bandpass, indicating a line FWHM of $>$338\,\kms. This suggests
that the line is significantly broader than in LBG-1 (which has a
total FWHM of 218$\pm$24\,\kms; Tab.~\ref{t1}). Ly$\alpha$-2 has a
$L_{\rm CII}$/$L_{\rm FIR}$ ratio of $>$5$\times$10$^{-4}$. Assuming
that at least half the [C{\scriptsize II}] line in Ly$\alpha$-2 is
covered by the bandpass, this suggests an at least
$\gtrsim$3--10$\times$ lower ratio than in LBG-1. These properties are
consistent with Ly$\alpha$-2 being a luminous, dusty starburst system,
perhaps not representative of $L^\star_{\rm UV}$ galaxies at
$z$$\sim$4.7, but less extreme than other dusty starbursts at high
redshift detected in [C{\scriptsize II}] previously.

The Ly$\alpha$-1 component in the BR\,1202--0725 system is not
detected in the far-infrared continuum, implying $L_{\rm CII}$/$L_{\rm
  FIR}$$>$5$\times$10$^{-4}$. This limit is by about an order of
magnitude lower than that in LBG-1. It thus remains unclear how its
$L_{\rm CII}$/$L_{\rm FIR}$ compares to normal, star-forming galaxies
nearby. The [C{\scriptsize II}] line in Ly$\alpha$-1 has a FWHM of
only 56$\pm$11\,\kms, corresponding to $\sim$4\% of the width of its
Ly-$\alpha$ line (Williams et al.\ \citeyear{wil14}). This also
corresponds to $\sim$60\% of the width of LBG-1a, i.e., the narrowest
component of LBG-1. As discussed by Carilli et
al.\ (\citeyear{car13}), it remains unclear if Ly$\alpha$-1 is a
physically distinct system, or a local maximum in a tidal ``bridge''
connecting the massive, far-infrared-luminous galaxies. In any case,
based on the existing constraints, the properties of Ly$\alpha$-1 and
LBG-1 appear to be quite dissimilar as well, but more detailed ALMA
data on the BR\,1202--0725 system will be required to further
investigate possible similarities.

It remains to be seen what role the environment plays in the evolution
of LBG-1. To address this issue in more detail, a first step will be a
complete study of the AzTEC-3 protocluster with ALMA (the current
study only covered the center of the region) and similar environments
to be discovered in the future. Equally importantly, sensitive studies
of ``blank fields'' in the [C{\scriptsize II}] emission line will be
necessary for an unbiased investigation of the [C{\scriptsize II}]
luminosity function (an approximate, but independent measure of the
atomic gas content of galaxies through cosmic times), and to properly
constrain the ``hidden'', dust-obscured part of the star formation
history of the universe through the detection of previously unknown
faint, dusty star-forming galaxies.  Such C$^+$ deep fields will be an
important complement to similar studies in CO and continuum emission
alone (studies of [C{\scriptsize II}] in continuum-preselected samples
will only yield a biased view of this issue; e.g., Swinbank et
al.\ \citeyear{swi12}).

When ALMA is completed in the coming months, it will be an ideal tool
for the most sensitive of these investigations. Given the relative
strength of the [C{\scriptsize II}] line, CCAT will be able to detect
[C{\scriptsize II}] emission over regions the size of the AzTEC-3
protocluster in a single shot using multi-object spectroscopy by the
end of the decade.  Ultimately, CCAT will also enable complementing,
large-area C$^+$ blank field studies that may cover regions as large
as the full COSMOS field to appreciable depth.

\acknowledgments

We thank the anonymous referee for a helpful and constructive report,
Jim Braatz for assistance with setting up the observations, Mark Lacy
for taking on the QA2 check of the data, and Alex Conley for help with
the SED fitting. The National Radio Astronomy Observatory is a
facility of the National Science Foundation operated under cooperative
agreement by Associated Universities, Inc.  This paper makes use of
the following ALMA data: ADS/JAO.ALMA\# 2011.0.00064.S. ALMA is a
partnership of ESO (representing its member states), NSF (USA) and
NINS (Japan), together with NRC (Canada) and NSC and ASIAA (Taiwan),
in cooperation with the Republic of Chile. The Joint ALMA Observatory
is operated by ESO, AUI/NRAO and NAOJ. FB and AK acknowledge support
from Collaborative Research Centre 956, sub-project A1, funded by the
Deutsche Forschungsgemeinschaft (DFG). VS acknowledges support by the
European Union's Seventh Frame-work program under grant agreement
337595 (ERC Starting Grant, ``CoSMass'').

\end{document}